\documentclass[preprint,aps,prd,showpacs,superscriptaddress,nofootinbib,tightenlines]{revtex4}

\usepackage{amsfonts}
\usepackage{mathrsfs}
\usepackage{graphicx}
\usepackage{amsmath}
\usepackage{amssymb}
\usepackage{bm}
\usepackage{bbm}
\usepackage{epsfig}
\usepackage{multirow}
\usepackage{float}
\usepackage{array}
\usepackage{tabularx}
\usepackage{xcolor}

\makeatletter
\def\hlinew#1{%
  \noalign{\ifnum0=`}\fi\hrule \@height #1 \futurelet
   \reserved@a\@xhline}
\usepackage{array}
\newcommand{\PreserveBackslash}[1]{\let\temp=\\#1\let\\=\temp}
\newcolumntype{C}[1]{>{\PreserveBackslash\centering}p{#1}}
\newcolumntype{R}[1]{>{\PreserveBackslash\raggedleft}p{#1}}
\newcolumntype{L}[1]{>{\PreserveBackslash\raggedright}p{#1}}
\usepackage{float}

\def\OMIT#1{}

\newcommand{\nn}{\nonumber}

\newcommand{\bqa}{\begin{eqnarray}}
\newcommand{\eqa}{\end{eqnarray}}

\newcommand{\jpsi}{J/\psi}

\begin{document}

%%%%%%%%%%%%%%%%%%%%%%%%%%%%%%%%%%%%%%%%%%
%Define Title, Author, Address, Preprint#

\title{Exclusive decay of $\bm{\Upsilon}$ into $\bm{J}\bm{/}\bm{\psi}\bm{+}\bm{\chi}_{\bm{c}\bm{0}
\bm{,}\bm{1}\bm{,}\bm{2}}$}

\author{Jia Xu\footnote{xuj@ihep.ac.cn}}
\affiliation{Institute of High Energy Physics, Chinese Academy of
Sciences, Beijing 100049, China\vspace{0.2cm}}

\author{Hai-Rong Dong\footnote{donghr@ihep.ac.cn}}
\affiliation{Institute of High Energy Physics, Chinese Academy of
Sciences, Beijing 100049, China\vspace{0.2cm}}

\author{Feng Feng\footnote{fengf@ihep.ac.cn}}
\affiliation{Center for High Energy Physics, Peking University,
Beijing 100871, China\vspace{0.2cm}}

\author{Ying-Jia Gao\footnote{josiagyj@hotmail.com}}
\affiliation{Key Laboratory of Frontiers in Theoretical Physics,
Institute of Theoretical Physics, Chinese Academy of Sciences,
Beijing 100190, China\vspace{0.2cm}}

\author{Yu Jia\footnote{jiay@ihep.ac.cn}}
\affiliation{Institute of High Energy Physics,  Chinese Academy of
Sciences, Beijing 100049, China\vspace{0.2cm}}
\affiliation{Theoretical Physics Center for Science Facilities,
Institute of High Energy Physics,\\ Chinese Academy of Sciences,
Beijing 100049, China\vspace{0.2cm}}

\date{\today\\ \vspace{1cm} }

%%%%%%%%%%%%%%%%%%%%%%%%%%%%%%%%%%%%%%%%%%

%%%%%%%%%%%%%%%%%%%%%%%%%%%%%%%%%%%%%%%%%%
%Create the title page

\begin{abstract}

We study the $\Upsilon$ exclusive decay into double charmonium,
specifically, the $S$-wave charmonium $ J/\psi$ plus the $P$-wave
charmonium $\chi_{c0,1,2}$ in the NRQCD factorization framework.
Three distinct decay mechanisms, {\it i.e.}, the strong,
electromagnetic and radiative decay channels are included and their
interference effects are investigated. The decay processes
$\Upsilon(1S,2S,3S)\to J/\psi+\chi_{c1,0}$ are predicted to have the
branching fractions of order $10^{-6}$, which should be observed in
the prospective Super $B$ factory.
\\

%12.38.-t Quantum chromodynamics
% 12.38.Bx  Perturbative calculations
% 13.25.Gv Decays of J/¦×, ¦´, and other quarkonia

\end{abstract}

\maketitle

%%%%%%%%%%%%%%%%%%%%%%%%%%%%%%%%%%%%%%%%%%
%\tighten
\newpage
%%%%%%%%%%%%%%%%%%%%%%%%%%%%%%%%%%%%%%%%%%
%Main body of the paper

\section{Introduction}

Anyone who has ever browsed the Meson Summary Table in the biannual
review of particle data group will be impressed by the extremely
rich decay channels of the $B$, $D$, and $J/\psi$
mesons~\cite{Beringer:1900zz}. Unlike the heavy flavored mesons,
which can only decay via the weak interaction, the unflavored heavy
quarkonia, decay through the heavy quark-antiquark annihilation
initiated exclusively by the strong and electromagnetic
interactions. Although quite a few decay channels have been
established for the charmonia system over the past few decades, the
experimental information about the bottomonia decay is still very
sparse. Until very recently, the \textsc{Belle} Collaboration has
observed a few exclusive decay channels of $\Upsilon(1S,2S)$ into
light hadrons for the first time, {\it e.g.}, into the vector-tensor
states and the axial-vector-pseudoscalar states~\cite{Shen:2012iq}.

Because of the much more copious phase space opened at the
bottomonium energy level, the typical branching fraction for a given
hadronic decay mode of a bottomonium is greatly diluted with respect
to that of a charmonium. Thanks to the weaker strong coupling at the
bottom mass scale, the perturbative QCD is expected to work more
reliably for the hadronic bottomonium decay than for the charmonium.

An interesting class of hadronic decay processes of bottomonium is
into double charmonium, which may presumably be predicted with less
uncertainties than into the light hadrons. In the recent years, some
exclusive decay processes of bottomonium into double charmonia have
been intensively studied in the perturbative QCD framework, {\it
e.g.}, $\eta_b\to J/\psi
J/\psi$~\cite{Jia:2006rx,Gong:2008ue,Braguta:2009xu,Sun:2010qx},
$\chi_{b0,1,2} \to J/\psi
J/\psi$~\cite{Braguta:2005gw,Braguta:2009df,Zhang:2011ng,Sang:2011fw,Chen:2012ih},
and $\Upsilon\to J/\psi+\eta_c$~\cite{Irwin:1990fn,Jia:2007hy}.
These studies are largely inspired by the various double-charmonium
production processes in $e^+e^-$ annihilation, which were first
observed at the $B$ factories a decade
ago~\cite{Abe:2002rb,Abe:2004ww,Aubert:2005tj}. Triggered by the
disquieting discrepancy between data and theory, since then a great
number of theoretical studies have been conducted for the processes
$e^+e^-\to
J/\psi+\eta_c$~\cite{Braaten:2002fi,Liu:2002wq,Hagiwara:2003cw,Ma:2004qf,Bondar:2004sv,
Bodwin:2006dm,Zhang:2005cha,Gong:2007db, He:2007te, Bodwin:2007ga,
Braguta:2008tg,Brambilla:2010cs,Dong:2012xx}, and $e^+e^-\to
J/\psi+\chi_{c0,1,2}$~\cite{Zhang:2008gp,Wang:2011qg,Dong:2011fb}.

Besides the tremendous amount of data near the $\Upsilon(4S)$, the
\textsc{Belle} experiment to date has also collected about $102$
million $\Upsilon(1S)$ samples and $158$ million $\Upsilon(2S)$
samples. Therefore, it appears more promising to observe the
double-charmonium production from the $\Upsilon$ decay than from the
$C$-even bottomonia decay. In Ref.~\cite{Jia:2007hy}, the exclusive
decay of $\Upsilon$ to a vector-pseudoscalar charmonium states was
studied in the framework of nonrelativistic QCD (NRQCD)
factorization~\cite{Bodwin:1994jh}. The corresponding branching
fraction was estimated to be of order $10^{-6}$ and seems to have a
good chance to be observed at the Super $B$ factory. In this work,
we further investigate the $\Upsilon$ decay into the $ J/\psi$ plus
a spin-triplet $P$-wave charmonium $\chi_{cJ}$ ($J$=0,1,2). This
work should be considered as a sequel of
Ref.~\cite{Jia:2007hy}~\footnote{The main results of this paper have
already been reported in Ref.~\cite{Xu:2011:thesis}. Nevertheless,
some significant improvements have been made in the current work:
{\it i.e.}, some errors in calculating the three-gluon channel in
\cite{Xu:2011:thesis} have been corrected, and the contribution from
the two-gluon-one-photon channel is also included.}.

Although neither of these exclusive decay modes have been observed
yet, several upper bounds for $\Upsilon(nS)$ inclusive decay into
$J/\psi$ or $\chi_{cJ}$ have already been placed
experimentally~\cite{Beringer:1900zz}:
%-------------
\bqa
%-------------
 & & {\cal B}[\Upsilon(1S) \to J/\psi+X]= (6.5\pm 0.7)\times
10^{-4}\,,\quad {\cal B}[\Upsilon(1S) \to \chi_{c0}+X]< 5 \times
10^{-3}\,,
%-------------
\nn \\
%-------------
 & &{\cal B}[\Upsilon(1S) \to \chi_{c1}+X]= (2.3\pm 0.7)\times
10^{-4}\,,\quad {\cal B}[\Upsilon(1S) \to \chi_{c2}+X]= (3.4\pm
1.0)\times 10^{-4}\,
%-------------
\nn \\
%-------------
&&{\cal B}[\Upsilon(2S) \to J/\psi+X]< 6\times 10^{-3}\,,\quad {\cal
B}[\Upsilon(4S) \to J/\psi+X]< 1.9\times 10^{-4}\,.
%-------------
\label{inclusive:Psi:production:ups}
%-------------
\eqa
%-------------
It will be interesting to examine to which extent these upper bounds
are saturated by the predicted branching fractions for
$\Upsilon(nS)\to J/\psi+\chi_{c0,1,2}$.

As is well known, the hadronic decay of $\Upsilon$ can be
categorized into three distinct classes: $b\bar{b}$ either
annihilating into three gluons (strong decay), or single photon
(electromagnetic decay), or two-gluons and a photon (radiative
decay). On the experimental ground, the inclusive decay rates from
these three decay channels have been available long
ago~\cite{Beringer:1900zz}:
%-------------
\bqa {\cal B}[\Upsilon\to ggg]: {\cal B}[\Upsilon\to \gamma^*\to
X]:{\cal B}[\Upsilon\to gg\gamma] &= & 82.7\%:7.5\%:2.2\%,
%-------------
\label{Br:inclusive:ratio}
%-------------
\eqa
%-------------
where these three branching ratios sum up to $1-\sum{\cal
B}[\Upsilon\to l^+l^-]=92.5\% $, as they should~\footnote{We have
not included the contribution from the radiative transition
$\Upsilon\to \eta_b\gamma$, which has a completely negligible
branching ratio.}.

For the exclusive hadronic decay $\Upsilon\to J/\psi+\chi_{c0,1,2}$,
one may also be interested in ascertaining the relative strength and
the interference pattern among these different decay channels. This
sort of study has been conducted for the process $\Upsilon\to
J/\psi+\eta_{c}$~\cite{Jia:2007hy}. Note that there have lasted
constant experimental efforts to infer the relative phase between
the strong and electromagnetic amplitudes in exclusive $J/\psi$ and
$\psi'$ decays into two light
mesons~\cite{LopezCastro:1994xw,Baldini:1998en,Suzuki:1999nb,
Wang:2003hy,Yuan:2003hj,Dobbs:2006fj}. As we will see later, being a
consequence of $m_b > m_c \gg \Lambda_{\rm QCD}$, the relative
phases among three distinct decay channels in our processes stem
from the short-distance loop contribution, which can actually be
calculated in perturbation theory.

The rest of the paper is organized as follows.
%-----------------------------------------------
In Section~\ref{pol:decay:rate:hel:sel:rule}, we express the
polarized and unpolarized decay rates in terms of the helicity
amplitudes and briefly state the helicity selection rule.
%-----------------------------------------------
In Section~\ref{NRQCD:Calculation}, we conduct the lowest order (LO)
calculation for each independent helicity amplitude associated with
the decays $\Upsilon\to J/\psi+\chi_{c0,1,2}$, within the NRQCD
factorization approach. The contributions from three distinct decay
channels, {\it i.e.}, electromagnetic, strong, and radiative decay
channels, are all included, and the analytic expressions for each of
the helicity amplitudes are given.
%-----------------------------------------------
In Section~\ref{phenomenology},  we present our predictions of the
interference pattern among three distinct decay channels for
$\Upsilon\to J/\psi+\chi_{c0,1,2}$, and of the polarized and
unpolarized partial decay widths and the corresponding branching
fractions for $\Upsilon(1S,2S,3S)$ decays into
$J/\psi+\chi_{c0,1,2}$. We find that it appears quite promising for
the prospective Super $B$ experiment to observe these hadronic decay
processes.
%-----------------------------------------------
Finally we summarize in Section~\ref{summary}.
%-----------------------------------------------
In Appendix~\ref{helicity:projectors}, we list the explicit
expressions of the 10 helicity projectors that are used in
Section~\ref{NRQCD:Calculation}.
%-----------------------------------------------

%----------------------------------------------------------------------
\section{Polarized decay rates and helicity selection rule}
\label{pol:decay:rate:hel:sel:rule}
%----------------------------------------------------------------------

It is of some advantage to utilize the helicity amplitude
formalism~\cite{Jacob:1959at,Haber:1994pe} to analyze the hard
exclusive reactions, in particular for the decay process studied in
this work. From the experimental perspective, the helicity
amplitudes can in principle be accessed via measuring the angular
distributions of the decay products of $J/\psi$ and $\chi_{cJ}$
($J=0,1,2$), provided that the statistics is sufficient. From the
theoretical viewpoint, some essential dynamics underlying
perturbative QCD is clearly encoded in the helicity amplitude
analysis, which becomes rather obscured if one only looks at the
unpolarized reaction rates.

We will work in the $\Upsilon$ rest frame throughout this work.
Suppose the spin projection of the $\Upsilon$ along the $\hat{z}$
axis to be $S_z$ (The $\hat{z}$ axis, say, may be chosen as the beam
direction of the $e^-$ and $e^+$ collider which resonantly produce a
$\Upsilon$ meson). Let $\lambda$, $\tilde\lambda$ denote the
helicities carried by the outgoing $J/\psi$ and $\chi_{cJ}$,
respectively, and $\theta$ signify the angle between the $J/\psi$
momentum $\bf P$ and the $\hat{z}$ axis. The differential polarized
decay rate can be expressed as~\cite{Jacob:1959at,Haber:1994pe}
%----------------------------
\bqa
%----------------------------
{d\Gamma [\Upsilon(S_z)\to
J/\psi(\lambda)+\chi_{cJ}(\tilde{\lambda})]\over d\cos \theta} &=&
{|{\bf P}|\over 16\pi M^2_\Upsilon} \left|
d^1_{S_z,\lambda-\tilde{\lambda}}(\theta) \right|^2 \left|{\mathcal
A}^J_{\lambda,\tilde{\lambda}}\right|^2,
%----------------------------
\label{Upsilon:diff:polar:decay:rate}
%----------------------------
\eqa
%----------------------------
where ${\mathcal A}^{J}_{\lambda,\tilde{\lambda}}$ ($J=0,1,2$)
characterizes the corresponding helicity amplitude, which
encompasses all the nontrivial QCD dynamics. The angular
distribution is fully dictated by the quantum numbers $S_z$,
$\lambda$ and $\tilde{\lambda}$ through the Wigner rotation matrix
$d^j_{m,m'}(\theta)$. Note that angular momentum conservation
constrains that $|\lambda-\tilde{\lambda}|\le 1$. In
(\ref{Upsilon:diff:polar:decay:rate}), the magnitude of the
three-momentum carried by the $J/\psi$ (or $\chi_{cJ}$) is
determined by
%----------------------------
\bqa
%----------------------------
|{\bf P}|=
{\lambda^{1/2}(M^2_{\Upsilon},M^2_{J/\psi},M^2_{\chi_{cJ}}) \over 2
M_\Upsilon},
%----------------------------
\label{3:momentum:Kallen:function}
%----------------------------
\eqa
%----------------------------
where $\lambda(x,y,z)=x^2+y^2+z^2-2xy-2yz-2zx$.

Integrating (\ref{Upsilon:diff:polar:decay:rate}) over the polar
angle, and averaging over all three possible $\Upsilon$
polarizations, one finds the integrated rate of $\Upsilon$ decay
into $J/\psi+\chi_{cJ}$ in the helicity configuration
$(\lambda,\tilde{\lambda})$ to be
%----------------------------
\bqa
%----------------------------
& & \Gamma [\Upsilon \to J/\psi(\lambda)+\chi_{cJ}(\tilde{\lambda})]
= {|{\bf P}|\over 16\pi M^2_\Upsilon} \left|{\mathcal
A}^J_{\lambda,\tilde{\lambda}}\right|^2 \int^1_{-1}\! d \cos\theta
\, {1\over 3} \sum_{S_z} \left|
d^1_{S_z,\lambda-\tilde{\lambda}}(\theta) \right|^2
%----------------------------
 \nn\\
%----------------------------
 & & = {|{\bf P}|\over 24 \pi M^2_\Upsilon} \left|{\mathcal
A}^J_{\lambda,\tilde{\lambda}}\right|^2.
%----------------------------
\label{polar:diff:rate:section}
%----------------------------
\eqa
%----------------------------

Since this decay process can be initiated by the strong or
electromagnetic interactions, one can resort to the parity
invariance to reduce the number of independent helicity amplitudes:
%----------------------------
\bqa
%----------------------------
{\mathcal A}^J_{\lambda,\tilde{\lambda}}=(-1)^J {\mathcal
A}^J_{-\lambda,-\tilde{\lambda}}.
%----------------------------
\label{parity:trans:hel:ampl}
%----------------------------
\eqa
%----------------------------
As a consequence, the helicity channel $\Upsilon\to
J/\psi(0)+\chi_{c1}(0)$ is strictly forbidden.

Starting from (\ref{polar:diff:rate:section}), one readily obtains
the unpolarized decay rate by summing the contributions from all the
allowed helicity channels:
%----------------------------
\begin{subequations}
%----------------------------
\bqa
%----------------------------
& & \Gamma[\Upsilon\to J/\psi + \chi_{c0}] = {|{\bf P}|\over 24 \pi
M_\Upsilon^2} \left( \left|{\mathcal A}^0_{0,0}\right|^2+2
\left|{\mathcal A}^0_{1,0}\right|^2\right),
%----------------------------
\\
%----------------------------
& & \Gamma[\Upsilon\to J/\psi + \chi_{c1}] = {|{\bf P}|\over 24 \pi
M_\Upsilon^2} \left(2 \left|{\mathcal A}^1_{1,0}\right|^2+ 2
\left|{\mathcal A}^1_{0,1}\right|^2 +  2 \left|{\mathcal
A}^1_{1,1}\right|^2 \right),
%----------------------------
\\
%----------------------------
& & \Gamma[\Upsilon\to J/\psi + \chi_{c2}] = {|{\bf P}|\over 24 \pi
M_\Upsilon^2} \left(\left|{\mathcal A}^2_{0,0}\right|^2+ 2
\left|{\mathcal A}^2_{1,0}\right|^2 +  2 \left|{\mathcal
A}^2_{0,1}\right|^2 + 2 \left|{\mathcal A}^2_{1,1}\right|^2 +  2
\left | {\mathcal A}^2_{1,2}\right|^2 \right).
%---------------------------------
\eqa
%----------------------------
\label{unpol:decay:rate:Jpsi:chicJ}
%----------------------------
\end{subequations}
%----------------------------
There are two, three and five independent helicity amplitudes for
$\Upsilon\to J/\psi+\chi_{cJ}$ ($J=0,1,2$), respectively, as
enforced by the angular momentum conservation. We have also included
a factor of 2 to account for the parity-doublet contributions.

One important piece of physics underlying the hard exclusive
reactions is that each helicity amplitude possesses a definite
power-law scaling in the inverse power of large momentum transfer,
controlled by the {\it helicity selection rule}
(HSR)~\cite{Brodsky:1981kj}. At asymptotically large $m_b$, the
polarized decay rate in our process scales as~\cite{Braaten:2002fi}:
%-------------------
\bqa
%-------------------
{\Gamma[\Upsilon\to J/\psi(\lambda) + \chi_{cJ}(\tilde{\lambda})]
\over \Gamma[\Upsilon\to \mu^+\mu^-]} & \propto & v^8
\left(m_c^2\over m_b^2 \right)^{2+|\lambda + \tilde{\lambda}|},
%-------------------
\label{helicity:selection:rule}
%-------------------
\eqa
%-------------------
here $v$ denotes the characteristic velocity of the charm quark
inside a charmonium.

Equation (\ref{helicity:selection:rule}) implies that the helicity
state which possesses the slowest asymptotic decrease, is the one
that conserves the hadron helicities $|\lambda+\tilde{\lambda}|=0$.
In line with the angular momentum conservation, the only possible
configuration is $(\lambda, \tilde{\lambda})=(0,0)$. For each unit
of the violation of the helicity conservation, there is a further
suppression factor of $1/m_b^2$. In the limit $m_b\to \infty$,
perhaps only the $(0,0)$ helicity state is phenomenological
relevant. Note that in NRQCD factorization language, the charm quark
is also treated as heavy, and in fact its mass acts as the agent of
violating the hadron helicity conservation.

We note that, the power-law scaling specified in
(\ref{helicity:selection:rule}) is in general subject to the mild
modifications due to the $\ln (m_b^2/m_c^2)$ from the loop
contribution. This logarithmic scaling violation will be examined in
detail in Section~\ref{NRQCD:Calculation}.

\section{The Calculation of the Helicity Amplitudes in NRQCD factorization approach}
\label{NRQCD:Calculation}

The hard exclusive decay process $\Upsilon\to J/\psi+\chi_{cJ}$
($J=0,1,2$) is characterized by two hard scales set by the bottom
and charm quark masses. This process can proceed via three separate
channels: the $b\bar{b}$ pair first annihilates into a single
photon, or three gluons, or two gluons plus a photon, subsequently
the highly virtual photon/gluons transition into two $c\bar c$
pairs, which finally materialize into two fast-moving charmonium
states.

Two influential perturbative QCD approaches are legitimate to
describe such type of decay process, {\it i.e.}, the light-cone
approach~\cite{Lepage:1980fj,Chernyak:1983ej} which is based on
twist expansion, and the NRQCD factorization
approach~\cite{Bodwin:1994jh} that is based on the quark velocity
expansion. As was seen in Sec.~\ref{pol:decay:rate:hel:sel:rule},
most helicity channels associated with the process $\Upsilon\to
J/\psi+\chi_{c0,1,2}$ are of helicity-suppressed type. This feature
impairs the practical usefulness of the light-cone approach, since
the higher-twist light-cone distribution amplitudes of charmonia are
rather poorly understood at present.

On the other hand, the NRQCD factorization approach, which is based
upon a completely different expansion strategy, does not confront
any obstacle in dealing with helicity-flipped channels. In the past
two decades, this framework has been widely applied to numerous
quarkonium decay and production processes~\cite{Bodwin:1994jh}. In
contrast with the light-cone approach, the nonpertubative input
parameters in NRQCD factorization approach are {\it numbers} (local
NRQCD matrix elements, or wave functions at the origin) rather than
{\it functions} (light-cone distribution amplitudes). In this
regard, NRQCD approach seems to be more economic and predictive than
the light-cone approach.

In this work, we will investigate the process $\Upsilon\to
J/\psi+\chi_{c0,1,2}$ in the framework of NRQCD factorization,
incorporating aforementioned  three distinct decay
mechanisms~\footnote{Unlike the exclusive double-charmonium
production in $e^+e^-$ annihilation, where a factorization theorem
in NRQCD has been proved to all orders in
$\alpha_s$~\cite{Bodwin:2008nf}, there has not yet existed any
rigorous proof for the validity of NRQCD approach to $\Upsilon\to
J/\psi+\chi_{c0,1,2}$.}. We will be content with the lowest order
accuracy in both the velocity expansion and the strong-coupling
constant expansion. We are aware that our results may be subject to
considerable uncertainty from various sources, yet still hope the
predicted decay rates may capture the correct order of magnitude.

At the LO in the bottom and charm velocity, one can expedite the
NRQCD approach calculation by invoking the covariant projection
method~\cite{Braaten:2002fi}, {\it i.e.}, first calculate the
on-shell $T$-matrix for  $b \bar{b}(Q) \to c\bar{c}(P) +
c\bar{c}(\widetilde{P})$, then project each quark-antiquark pair
onto the intended spin-color and orbital angular momentum states. In
this case, all the involved nonpertubative quarkonium-to-vacuum
NRQCD matrix elements can be well approximated by three (the first
derivative of ) wave functions at the origin for the quarkonia
$\Upsilon$, $J/\psi$, and $\chi_{cJ}$: $R_{\Upsilon}(0)$,
$R_{J/\psi}(0)$, and $R_{\chi_{cJ}}'(0)$. Each of them can be either
obtained from the quark potential models, or calculated from lattice
simulation, or directly extracted from the quarkonia decay data.

The product of these three nonpertubative wave functions at the
origin ubiquitously enters each helicity amplitude. Thus it seems
convenient to define a {\it reduced} dimensionless helicity
amplitude, of which these nonperturbative factors are explicitly
pulled out. First, let us introduce a mass ratio variable:
%-------------------
\bqa
%-------------------
& & r \equiv { m_c^2 \over m_b^2}.
%-------------------
\eqa
%-------------------

The reduced helicity amplitude, dubbed
$a^J_{\lambda,\tilde{\lambda}}$, is related to the standard helicity
amplitude as follows:
%------------------------------------
\bqa\label{aem}
%------------------------------------
\mathcal{A}^{J}_{\lambda,\tilde{\lambda}} \equiv
-\sqrt{8\pi}\,N_c^{3\over 2}\,{ R_{\Upsilon}(0) R_{J/\psi}(0)
R_{\chi_{c}}'(0) \over \sqrt{m_b}\, m_c^4}\, r^{1+{1\over
2}|\lambda+\tilde{\lambda}|}\,
 a^{J}_{\lambda,\tilde{\lambda}},
%----------------------------
\label{def:reduced:hel:ampl:a}
%------------------------------------
\eqa
%------------------------------------
where $N_c=3$ denotes the number of colors. Note that the scaling
factor dictated by the HSR has been explicitly factored out, the
reduced amplitude $a^{J}_{\lambda,\tilde{\lambda}}$ is thereby
expected to scale with $r$ as ${\cal O}(r^0)$.

Inserting (\ref{def:reduced:hel:ampl:a}) back into
(\ref{polar:diff:rate:section}), one can reexpress the integrated
polarized decay rate as
%----------------------------
\bqa
%----------------------------
\Gamma [\Upsilon \to J/\psi(\lambda)+\chi_{cJ}(\tilde{\lambda})] & &
=N_c^3 R_{\Upsilon}^2(0)R_{J/\psi}^2(0) R_{\chi_{cJ}}^{\prime\,2}(0)
{|{\bf P}|\over 3 M^2_\Upsilon m_b m_c^8}
r^{2+|\lambda+\tilde{\lambda}|}
%----------------------------
\nn \\
%----------------------------
& & \times
%----------------------------
\left|a
^J_{\gamma;\lambda,\tilde{\lambda}}+a^J_{3g;\lambda,\tilde{\lambda}}
+a^J_{\gamma gg;\lambda,\tilde{\lambda}} \right|^2,
%----------------------------
\label{polar:decay:rate:reduced:hel:ampl}
%----------------------------
\eqa
%----------------------------
for each helicity channel. The subscripts $\gamma$, $3g$, and
$\gamma gg$ emphasize the decay channel with which the reduce
amplitude is affiliated. Obviously, it is of interest to ascertain
the relative strength and phase among these different types of
amplitudes.

In the remainder of this section, we will present the analytic
expressions of the reduced helicity amplitudes associated with each
decay channel.

\subsection{Single-photon channel}
\label{sec:single-photon}

\begin{figure}[tb]
\begin{center}
\includegraphics[scale=0.6]{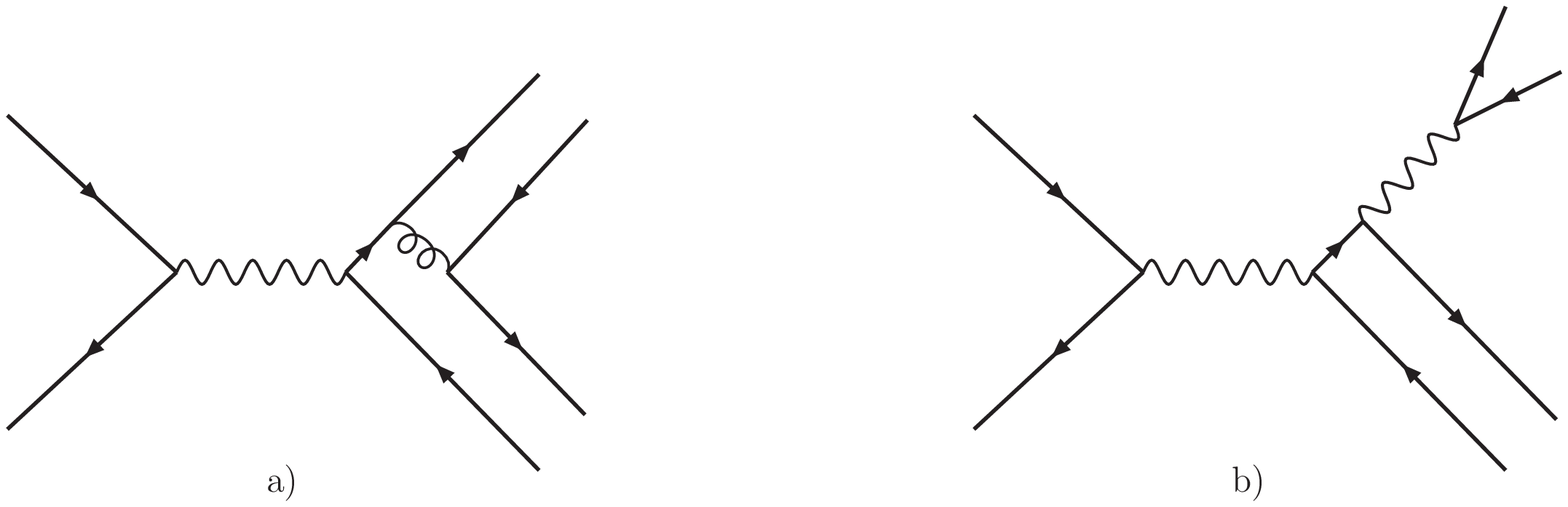}
\caption{Two representative lowest-order diagrams that contribute to
$\Upsilon \to \gamma^* \to J/\psi+\chi_{cJ}$.  There are totally
four diagrams in class a) and two in class b).
\label{feynman:diag:gamma}}

\end{center}
\end{figure}

We start by considering the decay channel $\Upsilon \to \gamma^* \to
J/\psi+\chi_{cJ}$, with some typical LO diagrams shown in
Fig.~\ref{feynman:diag:gamma}. This process is very similar to the
continuum $J/\psi+\chi_{cJ}$ production in $e^+e^-$
annihilation~\cite{Braaten:2002fi}.

After obtaining the decay amplitude ${\cal A}^J_\gamma$ in NRQCD
factorization, one can employ the helicity projectors enumerated in
Appendix~\ref{helicity:projectors} to project out 10 corresponding
helicity amplitudes. It is straightforward to follow equation
(\ref{def:reduced:hel:ampl:a}) to read off the reduced helicity
amplitude in the single-photon channel:
%------------------------------------
\bqa
%------------------------------------
& & a^ {J}_{\gamma;\lambda,\tilde{\lambda}}= {N_c^2-1 \over N^2_c}
e_b e_c\alpha \alpha_s\,
  c^{J}_{\lambda,\tilde{\lambda}}(r),
%------------------------------------
\label{channel:gamma:red:hel:ampl}
%------------------------------------
\eqa
%------------------------------------
where $e_b=-{1\over 3}$ and $e_c= {2\over 3}$ are the electric
charges of the $b$ and $c$ quarks, $\alpha$ and $\alpha_s$ are the
QED and QCD coupling constants, respectively. The coefficient
functions $c^{J}_{\lambda,\tilde{\lambda}}(r)$ read
%------------------------------------------------------------
\begin{subequations}
%------------------------------------------------------------
\bqa
%------------------------------------------------------------
& & c^{0}_{0,0}(r)= 1+10r-12 r^2+2r y \qquad c^{0}_{1,0}(r)= 9-14r-
{y\over 2}\left({1\over r}-6\right),
%------------------------------------------------------------
\\
%------------------------------------------------------------
& & c^{1}_{0,1}(r)= -\sqrt{6}\left[2-7r+{3\over 2}y\right]\qquad
c^{1}_{1,0}(r)= -\sqrt{6}\left[r+ {y\over 2}\left({1\over
r}-1\right)\right]
%------------------------------------------------------------
\nn \\
%------------------------------------------------------------
& &  c^{1}_{1,1}(r)= -2\sqrt{6}\left[1-3r+{y\over 4}\left({1\over
r}+2\right)\right],
%------------------------------------------------------------
\label{c:function:gamma*:chic1}
\\
%------------------------------------------------------------
& & c^{2}_{0,0}(r)= \sqrt{2}[1-2r-12r^2+2r y]\qquad c^{2}_{0,1}(r)=
\sqrt{6}\left[1-5r+{y\over 2}\right]
%------------------------------------------------------------
\nn\\
%------------------------------------------------------------
& & c^{2}_{1,0}(r)= \sqrt{2} \left[3-11r-{y\over 2}\left({1\over
r}-3\right)\right]\qquad c^{2}_{1,1}(r) = 2\sqrt{6}\left[1-3r -
{y\over 4}\left({1\over r}-2\right)\right]
%------------------------------------------------------------
\nn \\
%------------------------------------------------------------
& & c^{2}_{1,2}(r)=  \sqrt{3}\left[2-{y\over r}\right].
%------------------------------------------------------------
\eqa
%------------------------------------------------------------
\end{subequations}
%------------------------------------------------------------
where $y \equiv -\alpha/\alpha_s$. The $y$-dependent terms
characterize the photon fragmentation contributions as depicted in
Fig.~\ref{feynman:diag:gamma}b), which are often accompanied by an
enhancement factor $1/r$ for the transversely-polarized $J/\psi$.

Barring the pure QED fragmentation contributions, these 10
coefficient functions agree, up to an immaterial phase, with those
associated with the process $e^+e^-\to
J/\psi+\chi_{c0,1,2}$~\cite{Dong:2011fb}~\footnote{We take this
opportunity to point out a typo in equation (8c) in
Ref.~\cite{Dong:2011fb}, where $c^2_{1,0}(r)$ was erroneously typed
as $\sqrt{2}(11-3r)$.}. It is interesting to mention that, for some
accidental reason, the QCD part of the single-photon $J/\psi(\pm
1)+\chi_{c1}(0)$ amplitude receives an extra suppression factor than
implied from HSR.

\subsection{Three-gluon channel}
\label{subsect:three:gluon:channel}

\begin{figure}[tb]
\begin{center}
\includegraphics[scale=0.6]{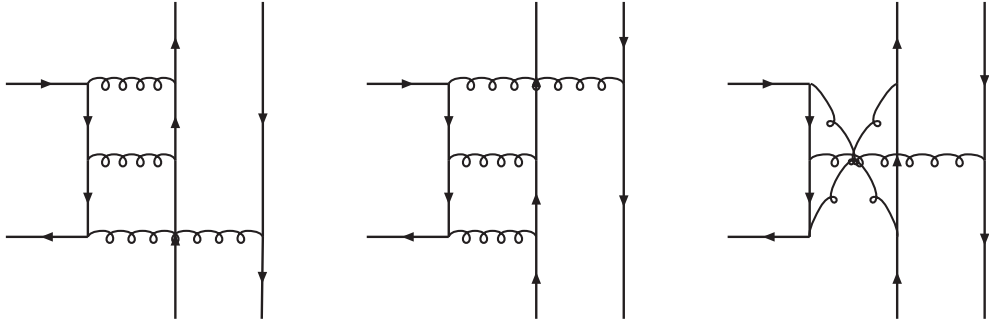}
\caption{Some representative lowest-order diagrams that contribute
to $\Upsilon \to 3g\to J/\psi+\chi_{cJ}$.
\label{feynman:diag:3g}}
\end{center}
\end{figure}

Next we turn to the strong decay channel $\Upsilon \to 3g\to
J/\psi+\chi_{cJ}$, which supposedly makes the most significant
contribution. Some of the representative LO diagrams have been
illustrated in Fig.~\ref{feynman:diag:3g}. In contrast with the
single-photon channel, this channel first starts at the one-loop
level. The charge conjugation invariance guarantees that one needs
only retain those diagrams with ``Abelian" gluon topology as shown
in Fig.~\ref{feynman:diag:3g}.

After obtaining the decay amplitude ${\cal A}^J_{3g}$ using the
covariant projection technique, and prior to performing the loop
integration, we apply those helicity projection operators given in
Appendix~\ref{helicity:projectors} to project out 10 corresponding
helicity amplitudes. This operation brings forth great
simplification, because all the polarization vectors (tensors) of
$\Upsilon$, $J/\psi$ and $\chi_{cJ}$ have been eliminated from the
integrand, and the numerators in loop integrals now become Lorentz
scalars comprised entirely of the external and loop momenta.

It is then straightforward to utilize the partial fractioning
technique to reduce all the higher-point one-loop integrals into a
set of 2-point and 3-point scalar integrals. Most of the encountered
scalar integrals can be found in the appendix of
Ref.~\cite{Jia:2007hy}, whose correctness has been numerically
checked by the \textsc{Mathematica} package
\textsc{LoopTools}~\cite{Hahn:1998yk}. There also arise some
nonstandard 2- and 3-point scalar integrals, which contain
propagators with quadratic power due to the projection of the
$P$-wave state. All of their analytic expressions can be readily
worked out.

As a crosscheck, we also employ the \textsc{Mathematica} package
\textsc{FIRE}~\cite{Smirnov:2008iw} and the code
\textsc{Apart}~\cite{Feng:2012iq} to perform an independent
calculation. Thanks to the integration-by-part (IBP) algorithm built
in \textsc{FIRE}, it turns out that all the required master
integrals (MIs) become just the conventional 2-point and 3-point
scalar integrals as given in \cite{Jia:2007hy}. The final results
generated by this more automatic approach exactly coincides with
those obtained from the partial-fractioning method.

As a third consistency check, the calculation is redone by
exchanging the order between helicity projection and loop
integration. That is, we first utilize the programs \textsc{Apart}
and \textsc{FIRE} at the amplitude level, which are more cumbersome
and time-consuming, yet still technically feasible. Once the
IR-finite $T$-matrices are obtained, we then project out each
intended helicity amplitudes at the very end. We again find the
exact agreement with the previous two methods. This calculation can
be viewed as a strong support for the validity of the 4-dimensional
helicity projectors given in Appendix~\ref{helicity:projectors}.

Each individual diagram in Fig.~\ref{feynman:diag:3g}, being
ultraviolet convergent, albeit contains logarithmic infrared
divergence. Dimensional regularization (DR) is adopted to regularize
those IR singularities. Upon summing all the diagrams, the ultimate
expression for each helicity amplitude becomes IR finite, which
endorses the validity of NRQCD factorization approach for these
exclusive $\Upsilon$ decay processes.

Following (\ref{def:reduced:hel:ampl:a}), we express the reduced
helicity amplitude in the three-gluon channel as
%------------------------------------
\bqa
%------------------------------------
& & a^ {J}_{3g;\lambda,\tilde{\lambda}}=
{(N_c^2-1)(N_c^2-4) \over N^4_c} {  \alpha_s^3 \over 8\pi}
{m_b^2\over m_b^2-4m_c^2}
  f^{J}_{\lambda,\tilde{\lambda}}(r).
%------------------------------------
\label{channel:3g:red:hel:ampl}
%------------------------------------
\eqa
%------------------------------------
The color factor reflects the fact the three ``Abelian" gluons in
Fig.~\ref{feynman:diag:3g} must bear an odd $C$-parity, thus
proportional to $d^{abc} d^{abc}= (N_c^2-1)(N_c^2-4)/ N_c$, where
$d^{abc}$ denotes the totally symmetric structure constants of
$SU(N_c)$ group.

\begin{figure}[tbh]
\includegraphics[width=0.3\textwidth]{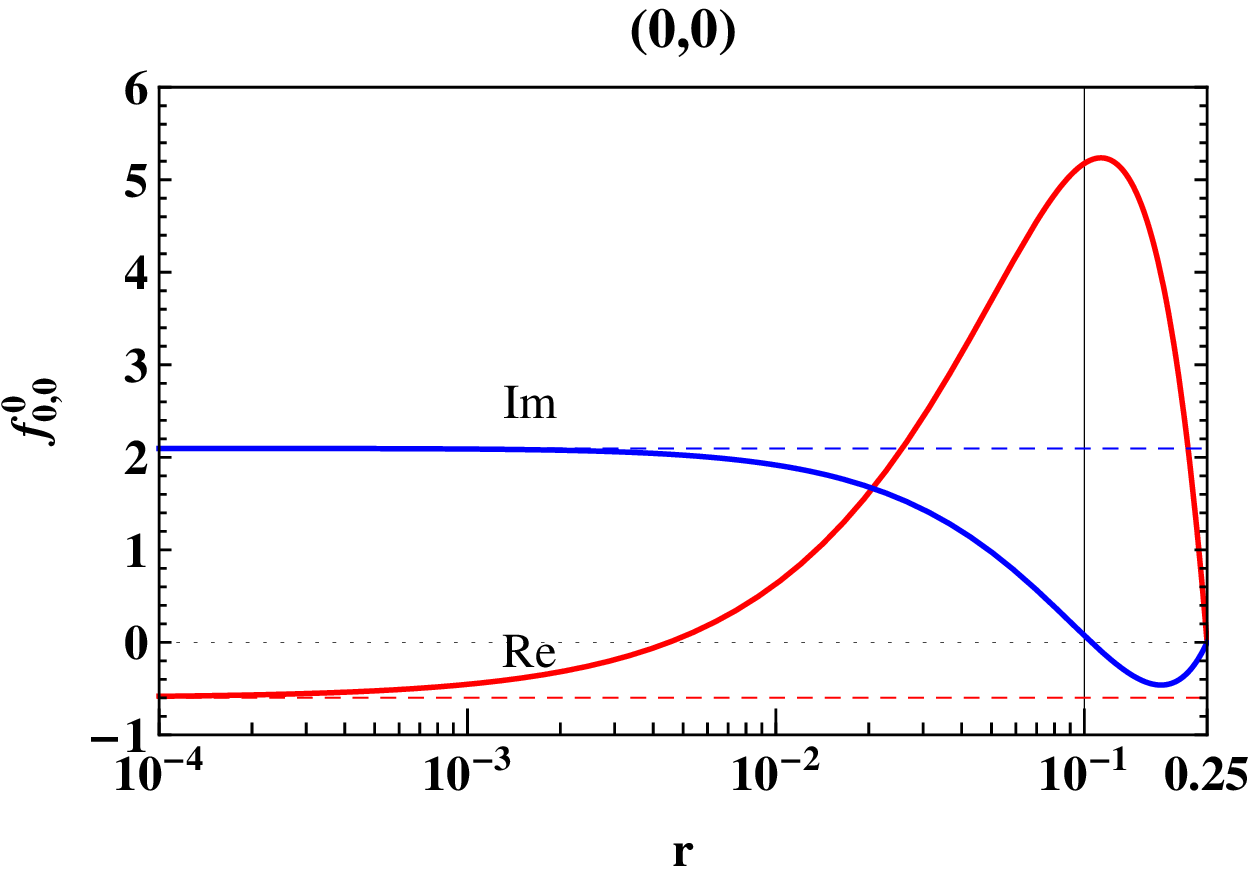}\ \ \
\includegraphics[width=0.3\textwidth]{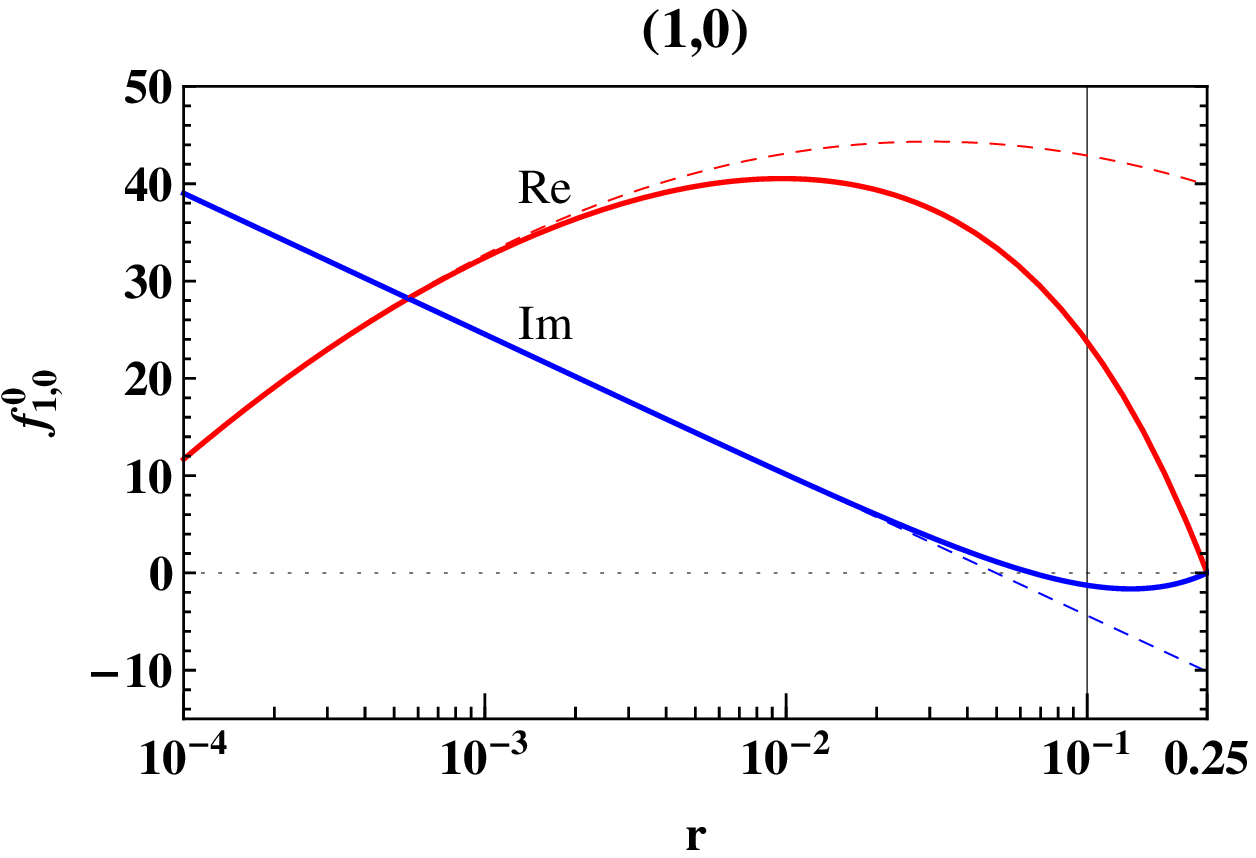}
\caption{Real and imaginary parts of
$f^0_{\lambda,\tilde{\lambda}}(r)$. The solid curves correspond to
the exact results, and the dashed curves represent the asymptotic
ones taken from (\ref{F:J:asym:expression}a) and
(\ref{F:J:asym:expression}b). The vertical mark is placed at the
phenomenologically relevant point $r=0.10$.
%--------------------
\label{plot:3g:F:0} }
%--------------------
\end{figure}

\begin{figure}[tbh]
%\raggedright
\includegraphics[width=0.3\textwidth]{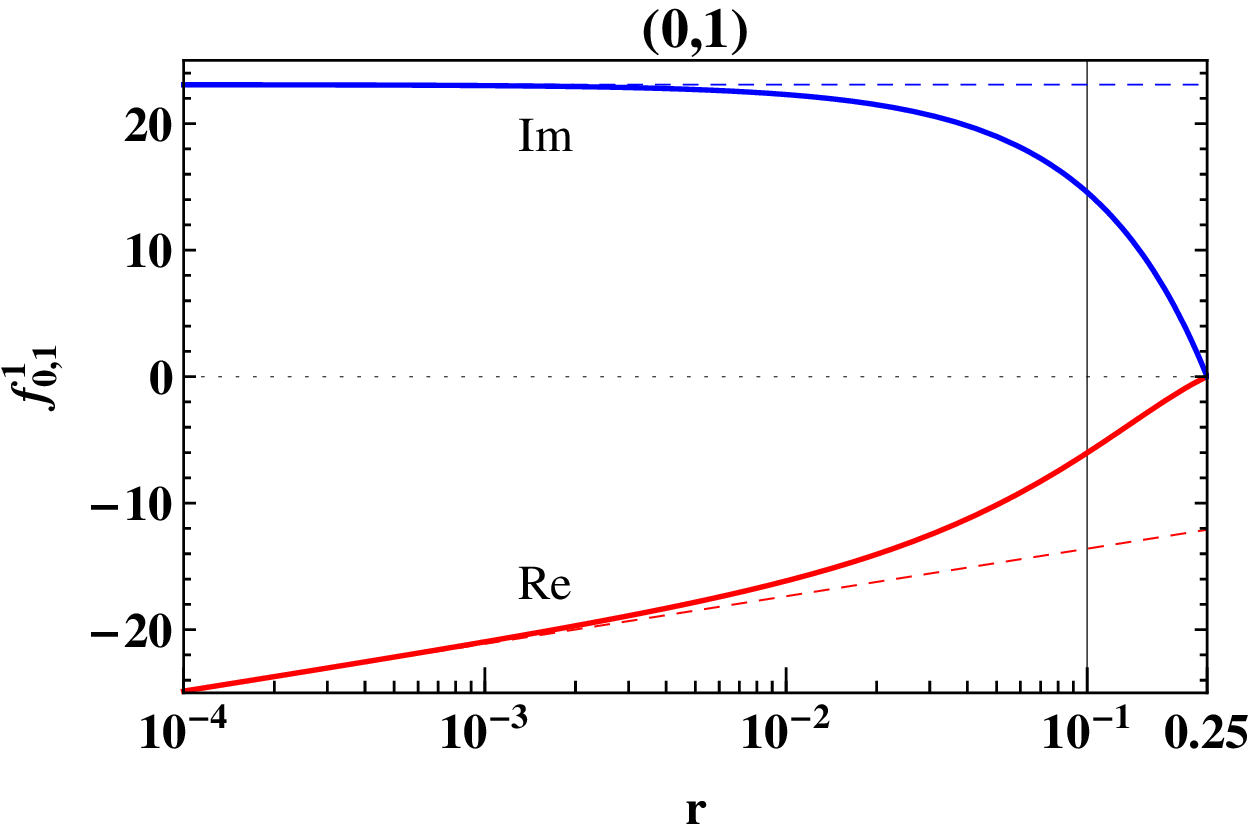}\ \ \
\includegraphics[width=0.3\textwidth]{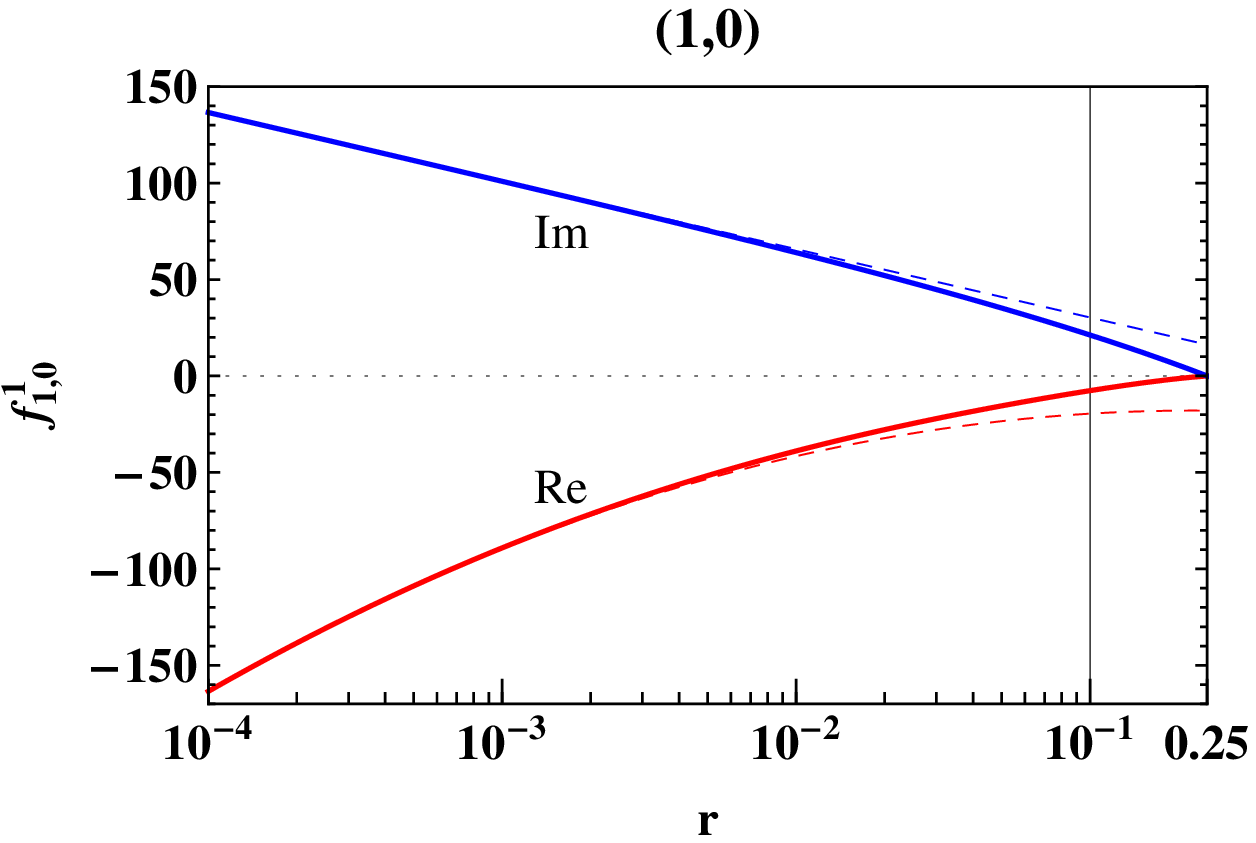}\ \ \
\includegraphics[width=0.3\textwidth]{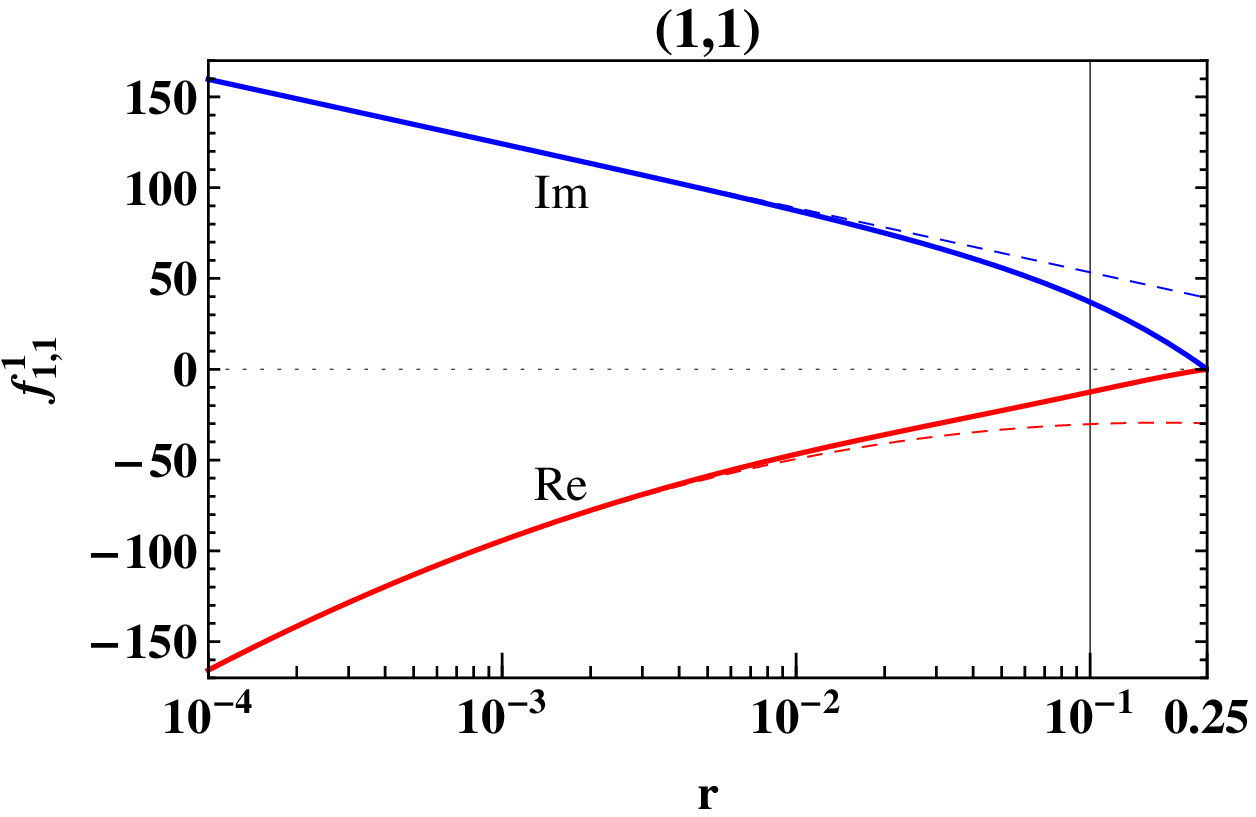}
\caption{Real and imaginary parts of
$f^1_{\lambda,\tilde{\lambda}}(r)$. The solid curves correspond to
the exact results, and the dashed curves represent the asymptotic
ones taken from (\ref{F:J:asym:expression}c) to
(\ref{F:J:asym:expression}e).
%--------------------
\label{plot:3g:F:1}}
%--------------------
\end{figure}

\begin{figure}[tbh]
%\raggedright
\includegraphics[width=0.3\textwidth]{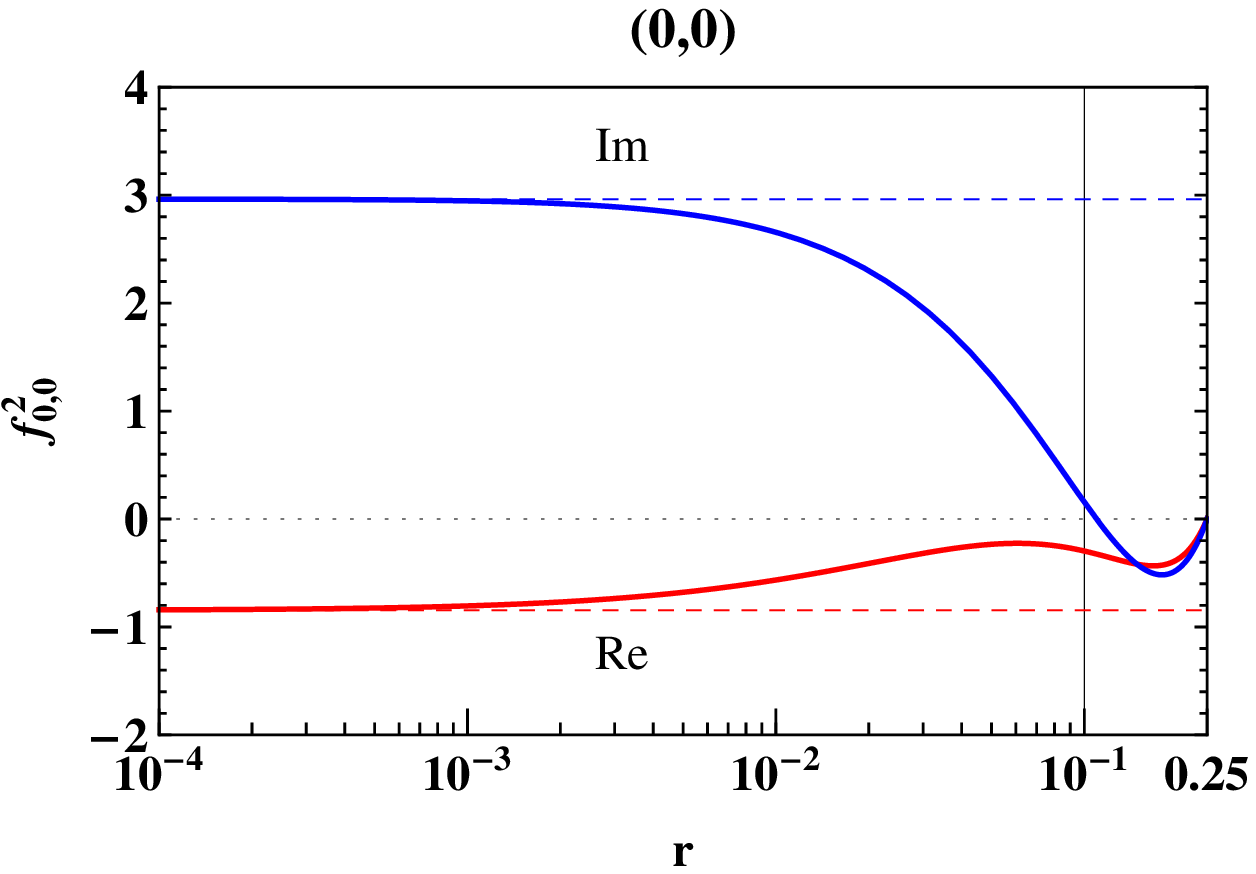}\ \ \
\includegraphics[width=0.3\textwidth]{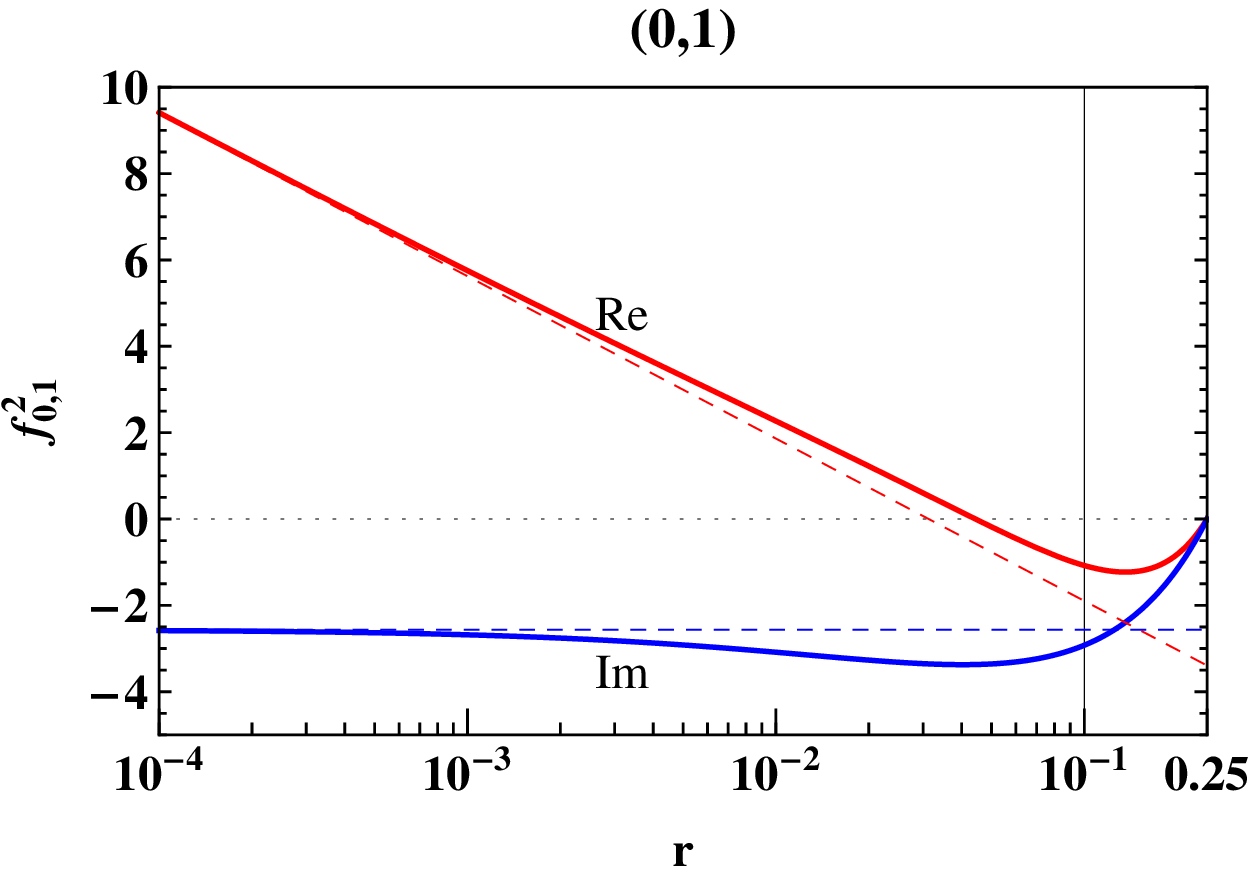}\ \ \
\includegraphics[width=0.3\textwidth]{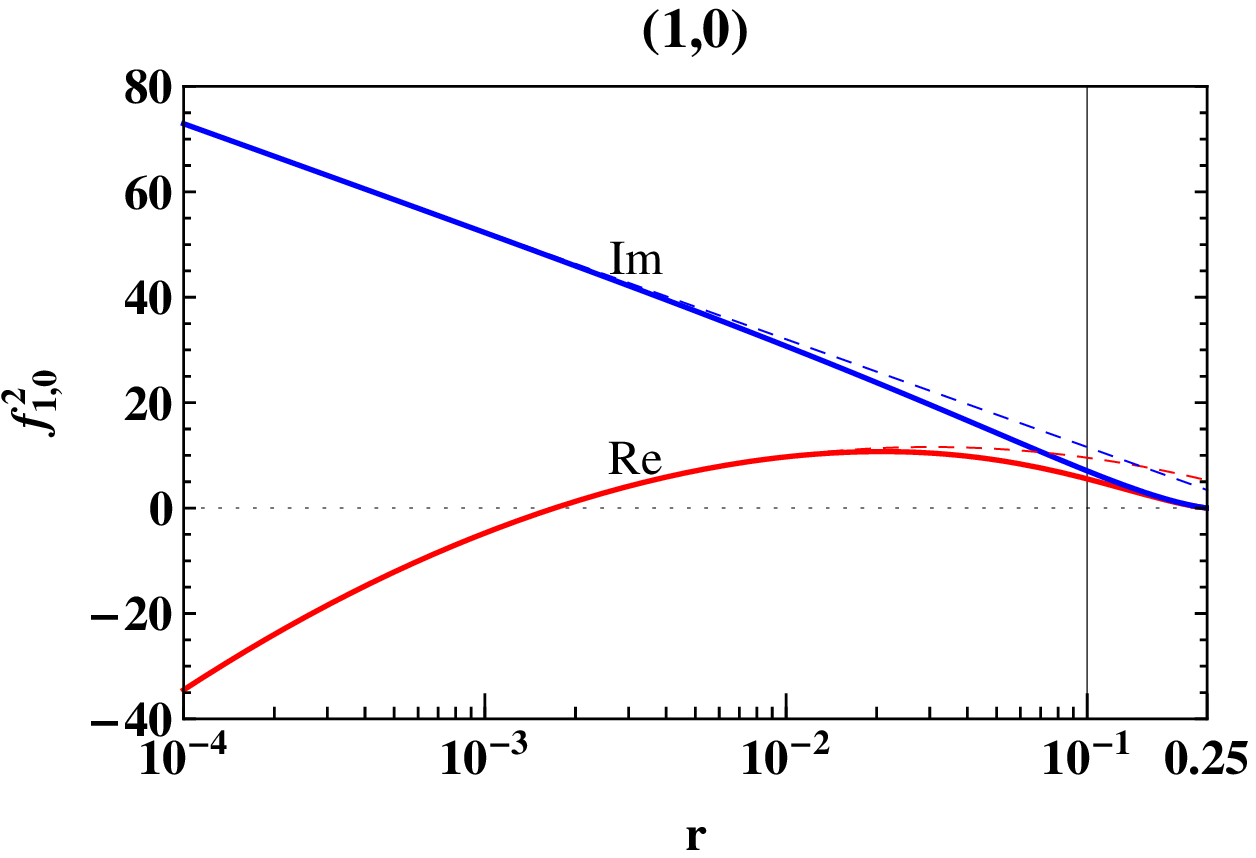}\ \ \
\includegraphics[width=0.3\textwidth]{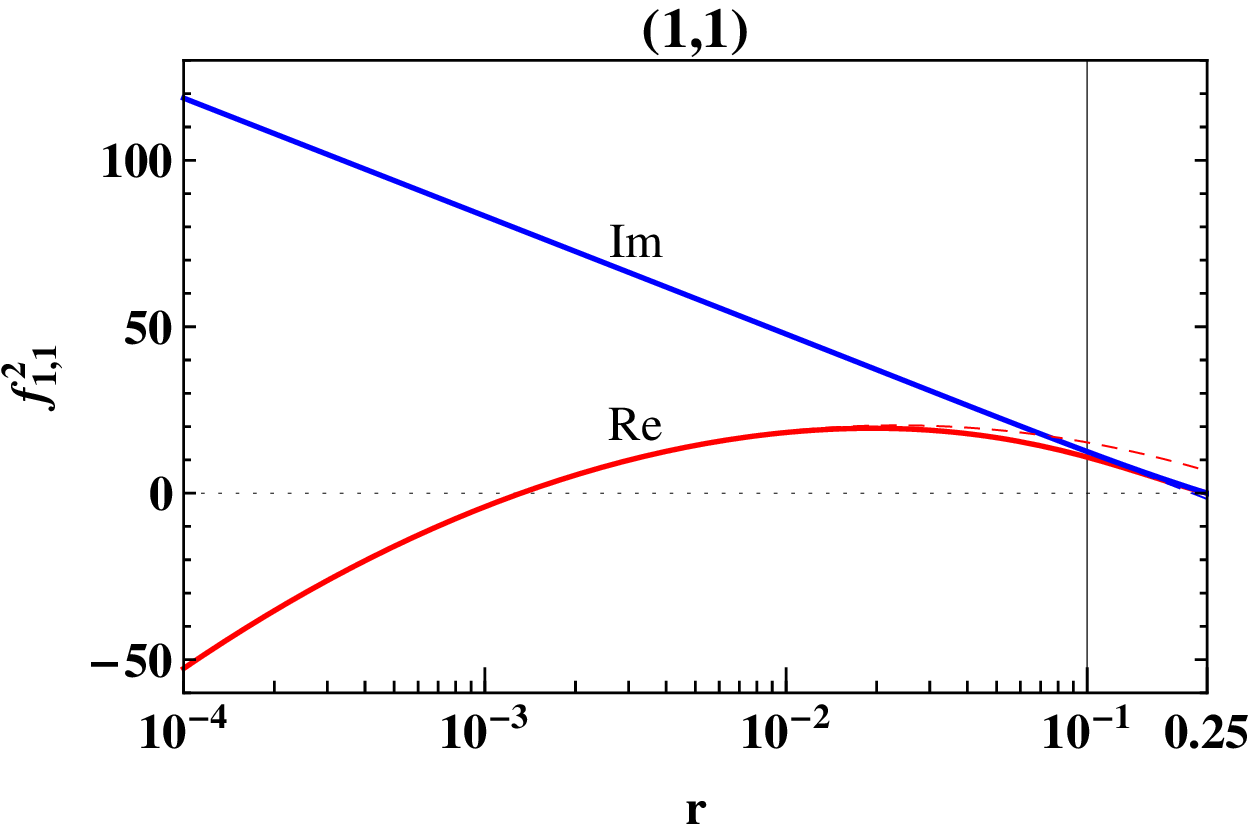}\ \ \
\includegraphics[width=0.3\textwidth]{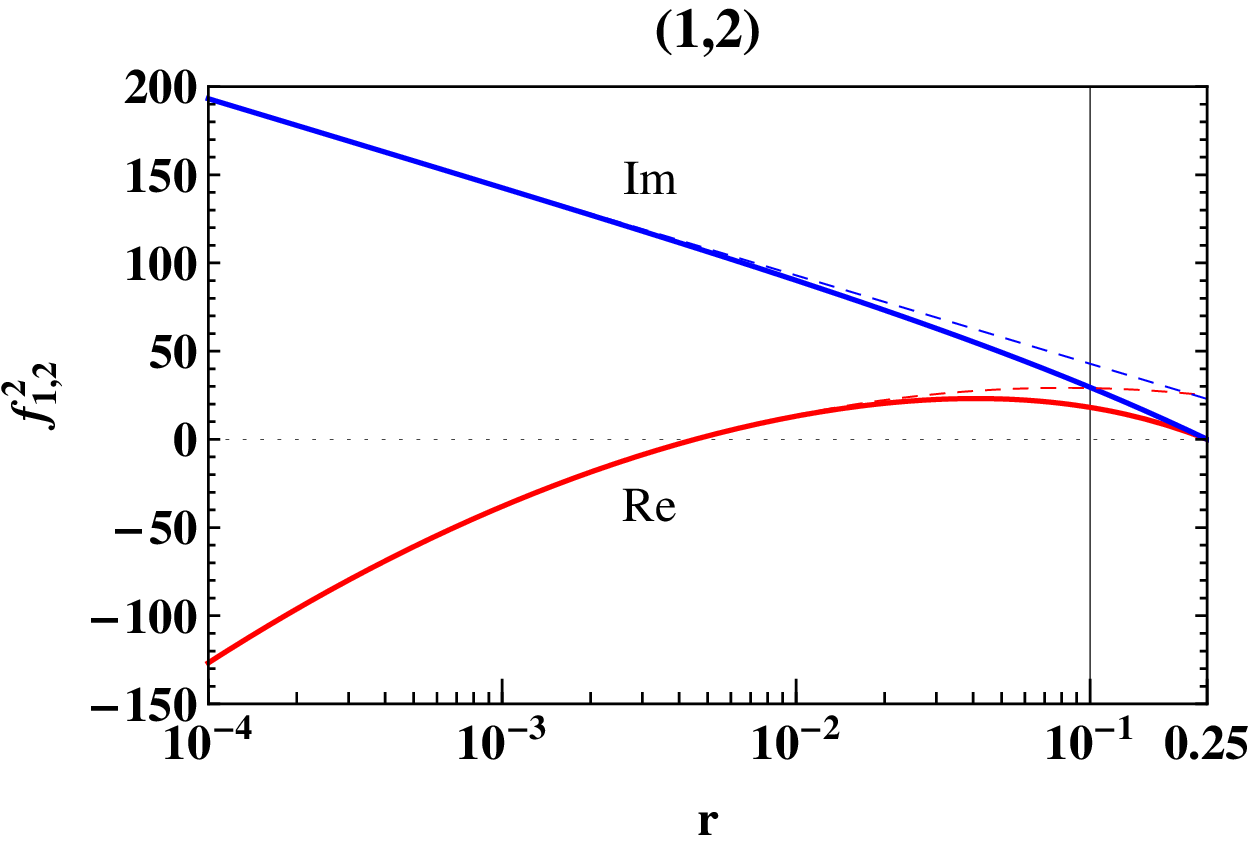}
\caption{Real and imaginary parts of
$f^2_{\lambda,\tilde{\lambda}}(r)$. The solid curves correspond to
the exact results, and the dashed curves represent the asymptotic
ones taken from (\ref{F:J:asym:expression}f) to
(\ref{F:J:asym:expression}j).
%--------------------
\label{plot:3g:F:2}}
%--------------------
\end{figure}

All the loop effects are encapsulated in the complex-valued
dimensionless functions $ f^{J}_{\lambda,\tilde{\lambda}}(r)$. Their
full expressions are somewhat lengthy, so will not be reproduced
here~\footnote{In our previous calculation reported in
Ref.~\cite{Xu:2011:thesis}, prior to performing the loop
integration, we erroneously carried out the Dirac trace in 4
spacetime dimensions. This is an unfortunate mistake which
contradicts the spirit of DR. In the current work, we take all the
Lorentz vectors (both loop and external momenta) as the
$D=4-2\epsilon$ dimensional objects when calculating the Dirac
trace. However, we would like to stress that, the helicity
projectors listed in Appendix~\ref{helicity:projectors}, which are
derived by simply assuming $D=4$, are still applicable in this
situation. That is because the ultimate amplitudes are UV, IR
finite, so it does not matter whether the external momenta are taken
as $4-2\epsilon$- or $4$-dimensional in the intermediate steps.
Finally, we note that the analytic expressions of the various
reduced helicity amplitudes in the three-gluon channel for
$\Upsilon\to J/\psi+\chi_{c0,1,2}$ markedly differ from those given
in Ref.~\cite{Xu:2011:thesis}.}. On the other hand, the profiles of
these functions over a wide range of $r$ are explicitly shown in
Figs.~\ref{plot:3g:F:0}, \ref{plot:3g:F:1}, \ref{plot:3g:F:2}.

%--------------------
%--------------------

%--------------------
%--------------------

%--------------------
%--------------------

Theoretically, it is interesting to ascertain the asymptotic
behaviors of these reduced helicity amplitudes in the limit $m_b\gg
m_c$. As mentioned before, we anticipate to see the logarithm
scaling violation to the naive power-law scaling given in
(\ref{helicity:selection:rule}). The asymptotic expressions of the
$f^J_{\lambda,\tilde{\lambda}}$ functions read ($J=0,1,2$):
%-------------------
\begin{subequations}
%-------------------
\bqa
%-------------------
f^0_{0,0}(r)&=&  -4 +{2\pi\over 3\sqrt{3}} + {2\pi^2\over 9} + {2
\pi i \over 3} +{\cal O}(r\ln^2{r}),
%-------------------
\\
%-------------------
f^0_{1,0}(r) &=& -\ln^2 r -7\ln{r} -{\pi^2\over 9}
+{25\pi\over\sqrt{3}} - {73\over 6} - 2 \pi i (\ln{r}+3)+ {\cal
O}(r\ln^2{r}),
%-------------------
\\
%-------------------
f^1_{0,1}(r) &=& \sqrt{6}\left( {2\over 3} \ln r + {\pi^2\over 3}-
{17\pi \over 2\sqrt{3}} + {73\over9} + 3 \pi i \right) +{\cal
O}(r\ln{r}),
%-------------------
\\
%-------------------
f^1_{1,0}(r) &=& \sqrt{6}\left[ -\ln^2 r -3\ln r + {\pi^2\over 9}-
{8\pi\over \sqrt{3}} + {23\over 6} - 2 \pi i \left(\ln{r}+{1\over
3}\right) \right]+ {\cal O}(r\ln^2{r}),
%-------------------
\\
%-------------------
f^1_{1,1}(r) &=& \sqrt{6}\left[ -\ln^2 r- {7\over 2} \ln
r-{8\pi^2\over 9}- {7\pi \over 3\sqrt{3}} - {25\over 12} - 2\pi i
\left( \ln{r}-{7\over 6}\right) \right] + {\cal O}(r\ln^2{r}),
%-------------------
\\
%-------------------
%-------------------
f^2_{0,0}(r)&=& \sqrt{2}\left( {2\pi^2\over 9} + {2\pi \over
3\sqrt{3}} -4 + {2\pi i \over 3} \right)+ {\cal O}(r\ln^2{r}),
%-------------------
\\
%-------------------
f^2_{0,1}(r)&=& \sqrt{6}\left( -{2 \over 3} \ln r + {5\pi^2\over 9}-
{7\pi\over 2\sqrt{3}}-{13\over 9} - {\pi i\over 3} \right) +{\cal
O}(r\ln^2{r}),
%-------------------
\\
%-------------------
f^2_{1,0}(r) &=& \sqrt{2}\left[-\ln^2 r -7\ln r + {11\pi^2\over 9} -
{11\pi\over \sqrt{3}} + {23\over 6} - 2\pi i(\ln{r}+1)\right]+{\cal
O}(r\ln^2{r}),
%-------------------
\\
%-------------------
f^2_{1,1}(r) &=& \sqrt{6}\left[ -\ln^2 r- {15 \over 2}\ln r
-{8\pi^2\over 9} +{25\pi\over 3\sqrt{3}} -{145\over 12} -2 \pi i
\left(\ln{r}+ {3\over 2}\right) \right] +{\cal O}(r\ln^2{r}),
%-------------------
\nn \\ \\
%-------------------
f^2_{1,2}(r)&=& \sqrt{3}\left[ -2\ln^2 r- 10\ln r +{10\pi^2\over 9}
-{20\pi\over \sqrt{3}} +{89\over 3} -4 \pi i \left(\ln{r}+ {1\over
3}\right) \right]+{\cal O}(r\ln^2{r}).\nn\\
%-------------------
\eqa \label{F:J:asym:expression}
%-------------------
\end{subequations}
%-------------------

For readers' convenience, all the asymptotic results of
$f^J_{\lambda,\tilde{\lambda}}(r)$ are also shown in
Figs.~\ref{plot:3g:F:0}, \ref{plot:3g:F:1}, \ref{plot:3g:F:2}, in
juxtapose with the corresponding exact results. We observe that for
most helicity configurations, the asymptotic results do not coincide
well with the exact ones at the phenomenologically relevant point
$r=m_c^2/m_b^2\approx 0.10$.

From (\ref{F:J:asym:expression}), one confirms that the scaling
violation is indeed of the logarithmic form. More interestingly, we
see that the occurrence of the double-logarithm $\ln^2 r$ is always
affiliated with the helicity-suppressed
($|\lambda+\tilde{\lambda}|>0$) decay channels. This is compatible
with the empirical pattern observed in
Ref.~\cite{Jia:2010fw,Dong:2011fb}. Indeed, such double logarithms
have previously also been observed in the helicity-suppressed
bottomonium exclusive decay processes, {\it e.g.} $\Upsilon\to
J/\psi+\eta_c$~\cite{Jia:2007hy}, $\eta_b\to J/\psi
J/\psi$~\cite{Gong:2008ue}. The light-cone approach is presumably
the proper tool to handle these process-dependent double logarithms.
Unfortunately, due to some longstanding problems, it remains as a
challenge to have a systematic control over these double logarithms
appearing in the NRQCD short-distance
coefficients~\cite{Jia:2010fw}.

Finally we mention one peculiarity associated with the helicity
channel $\Upsilon \rightarrow J/\psi(\lambda=\pm
1)+\chi_{c1}(\tilde{\lambda}=0)$. Recall that at LO in $\alpha_s$,
this helicity amplitude from the single-photon channel is suppressed
by an extra factor of $r$ with respect to the HSR, as can be seen in
(\ref{c:function:gamma*:chic1}). Nevertheless, from
(\ref{F:J:asym:expression}d), we find that this helicity amplitude
in the $3g$ channel just possesses the correct power-law scaling as
dictated by the HSR. This implies that the power suppression of the
LO single-photon amplitude is purely accidental.

\subsection{One-photon-two-gluon channel}

%--------------------
\begin{figure}[tb]
%--------------------
\begin{center}
%--------------------
\includegraphics[scale=0.6]{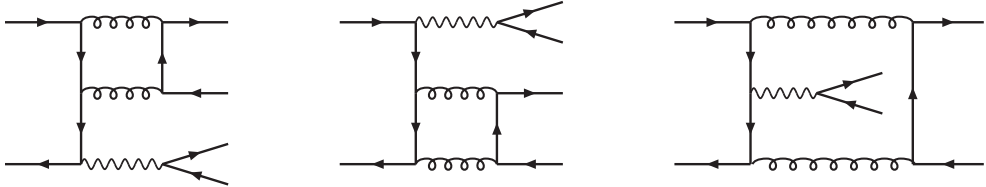}
\caption{Some representative lowest-order diagrams that contribute
to $\Upsilon \to \gamma gg\to J/\psi+\chi_{cJ}$.
%--------------------
\label{feynman:diag:gammagg} }
%--------------------
\end{center}
%--------------------
\end{figure}
%--------------------

From the ratios of the measured inclusive $\Upsilon$ decay rates in
three different channels, as listed in (\ref{Br:inclusive:ratio}),
an educated guess is that the exclusive decay channel $\Upsilon \to
\gamma gg \to J/\psi+\chi_{cJ}$ yields the least important
contribution. Nevertheless, for the sake of completeness, let us
finally assess the contribution from this radiative decay channel.

Similar to the strong decay channel, this radiative decay channel
also starts at the one-loop order. Some typical LO diagrams have
been illustrated in Fig.~\ref{feynman:diag:gammagg}. For simplicity,
and to be commensurate with the approximation adopted for the
single-photon channel in Sec.~\ref{sec:single-photon}, we only
retain those photon-fragmentation diagrams, whereas the neglected
diagrams are the identical as Fig.~\ref{feynman:diag:3g} except with
the gluon outside the loop replaced by the photon. We wish that
these fragmentation-type diagrams constitute the dominant
contributions, which is certainly the case for the
transversely-polarized $J/\psi$.

Following the steps outlined in
Sec.~\ref{subsect:three:gluon:channel}, one then projects out the 10
required helicity amplitudes, with all the polarization vectors
(tensors) of $\Upsilon$, $J/\psi$ and $\chi_{cJ}$ eliminated from
the loop integral. Nevertheless, it appears to be much more
difficult than in the three-gluon channel to employ the partial
fraction to simplify the encountered one-loop integrals.

Fortunately, the powerful \textsc{Mathematica} packages
\textsc{FIRE}~\cite{Smirnov:2008iw} and
\textsc{Apart}~\cite{Feng:2012iq} can still be successfully applied
to reduce the general higher-point tensor one-loop integrals into a
set of MIs. With the aid of the IBP algorithm built in
\textsc{FIRE}, all the involved MIs become just the standard 2-point
and 3-point scalar integrals. The analytic expressions of these
scalar integrals can be found in \cite{Jia:2007hy,Ellis:2007qk},
whose correctness have been numerically verified by using
\textsc{LoopTools}~\cite{Hahn:1998yk}.

Analogous to the three-gluon decay channel, each individual diagram
in Fig.~\ref{feynman:diag:gammagg} is UV finite but IR divergent.
After summing all the diagrams, the ultimate expression for each
helicity amplitude turns out to be completely IR finite.

In accordance with (\ref{def:reduced:hel:ampl:a}), the reduced
helicity amplitude in the radiative decay channel can be expressed
as
%------------------------------------
\bqa
%------------------------------------
& & a^ {J}_{\gamma gg;\lambda,\tilde{\lambda}}= {N_c^2-1 \over
N^2_c}\,{ e_b e_c \alpha \alpha_s^2 \over 4\pi}\,{m_b^2\over
m_b^2-4m_c^2}\,r^{-|\lambda|}
  g^{J}_{\lambda,\tilde{\lambda}}(r),
%------------------------------------
\label{channel:gammagg:red:hel:ampl}
\eqa
%------------------------------------
The inclusion of an extra factor $r^{-|\lambda|}$ is reminiscent of
the photon fragmentation mechanism: when the $J/\psi$ becomes
transversely polarized ($\lambda=\pm 1$), the corresponding helicity
amplitude would receive a $1/r$ enhancement with respect to the
nominal HSR.

\begin{figure}[htb]
\includegraphics[width=0.3\textwidth]{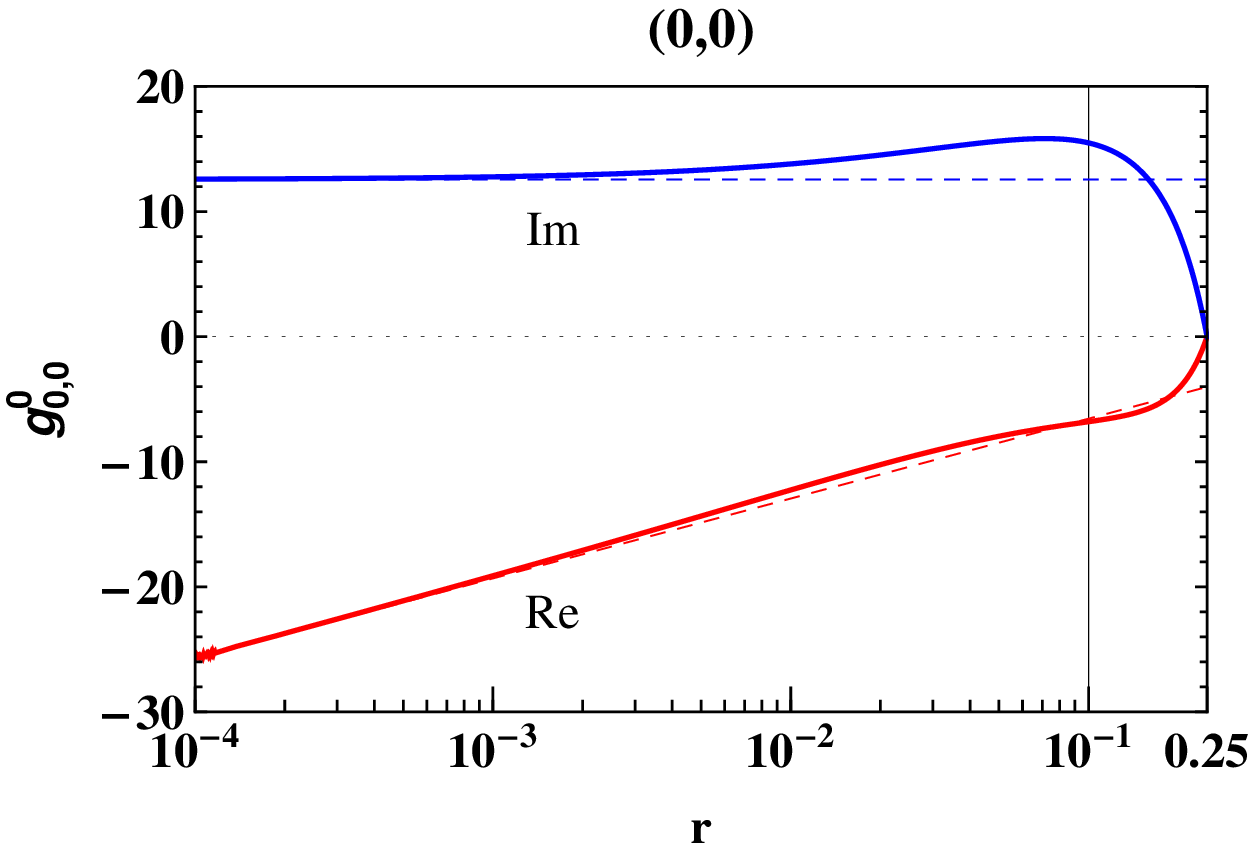}\ \ \
\includegraphics[width=0.3\textwidth]{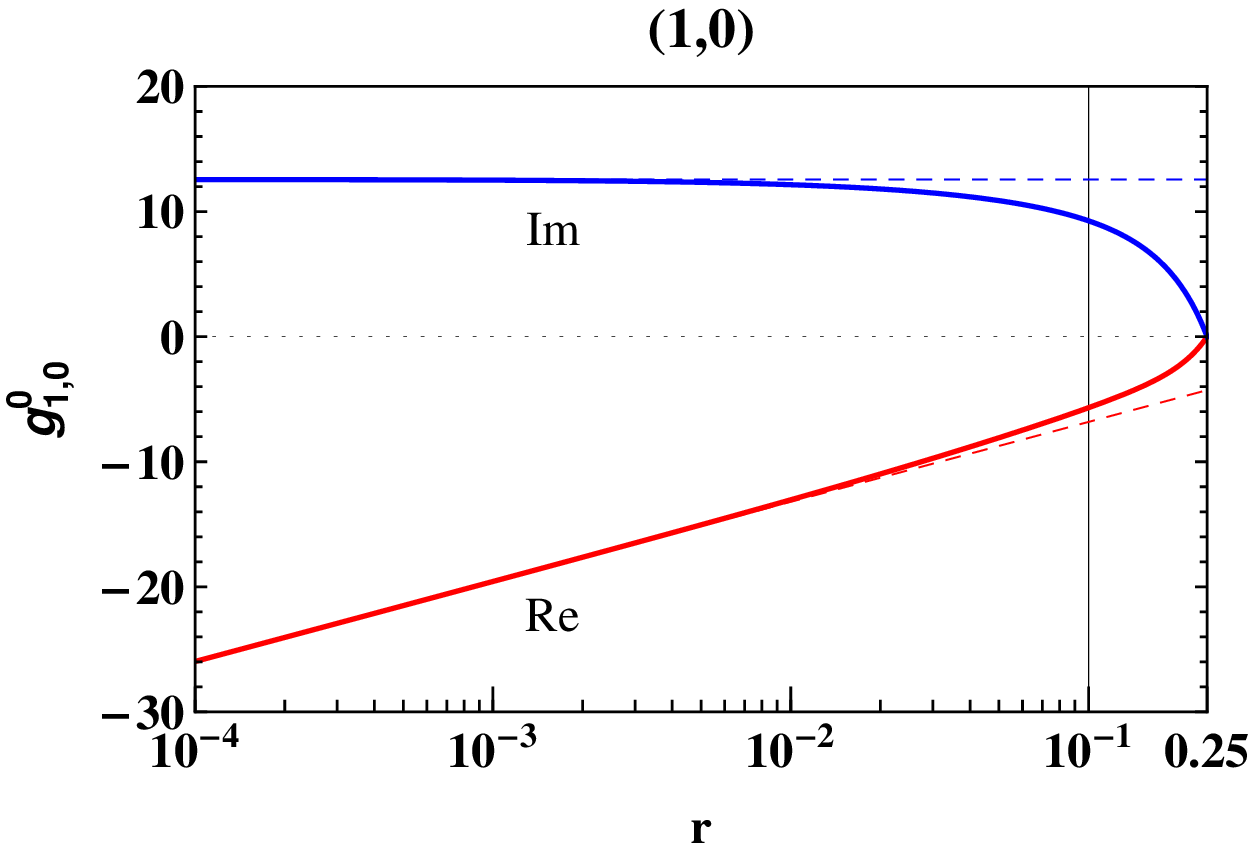}
\caption{Real and imaginary parts of
$g^0_{\lambda,\tilde{\lambda}}(r)$. The solid curves correspond to
the exact results, and the dashed curves represent the asymptotic
ones taken from (\ref{g:J:asym:expression}a) and
(\ref{g:J:asym:expression}b). The vertical mark is placed at the
phenomenologically relevant point $r=0.10$.
%--------------------
\label{plot:2g1gamma:g:0} }
%--------------------
\end{figure}

\begin{figure}[htb]
%\raggedright
\includegraphics[width=0.3\textwidth]{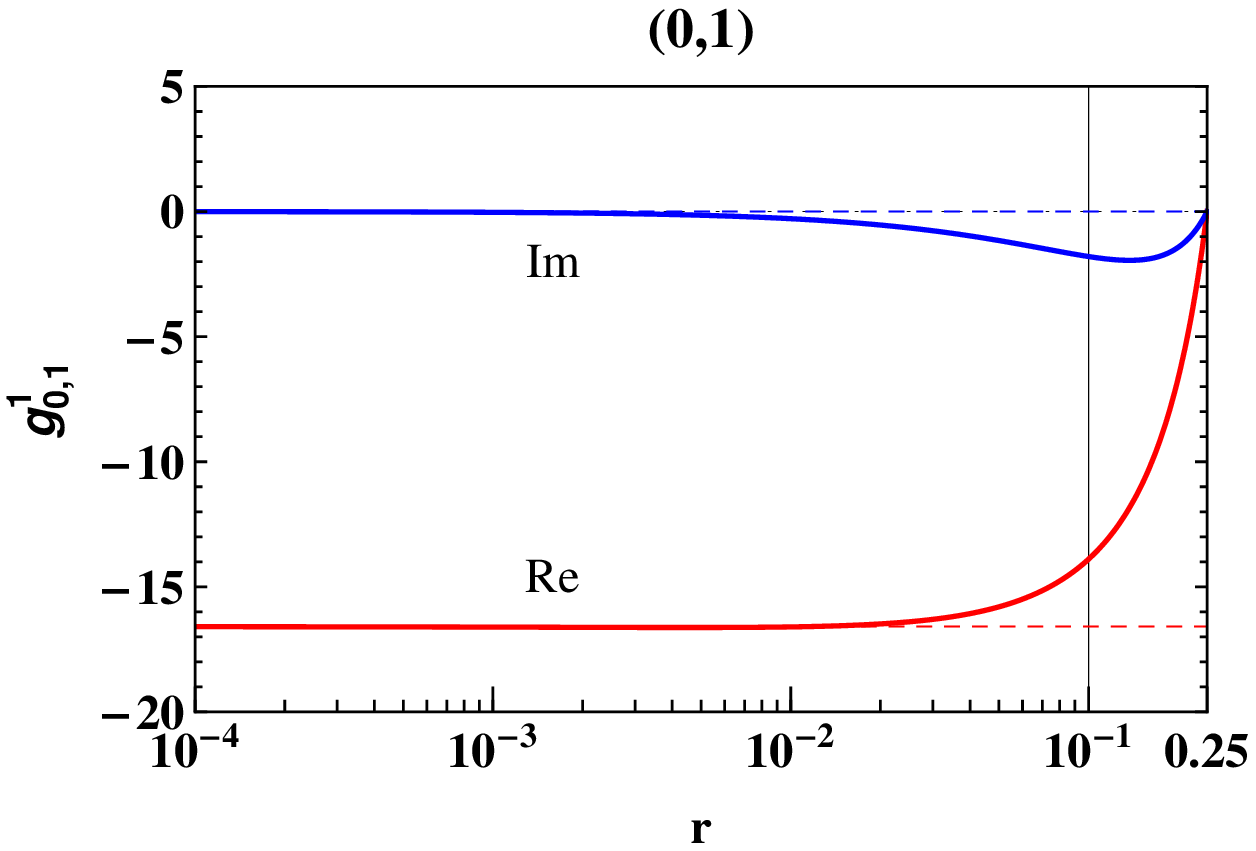}\ \ \
\includegraphics[width=0.3\textwidth]{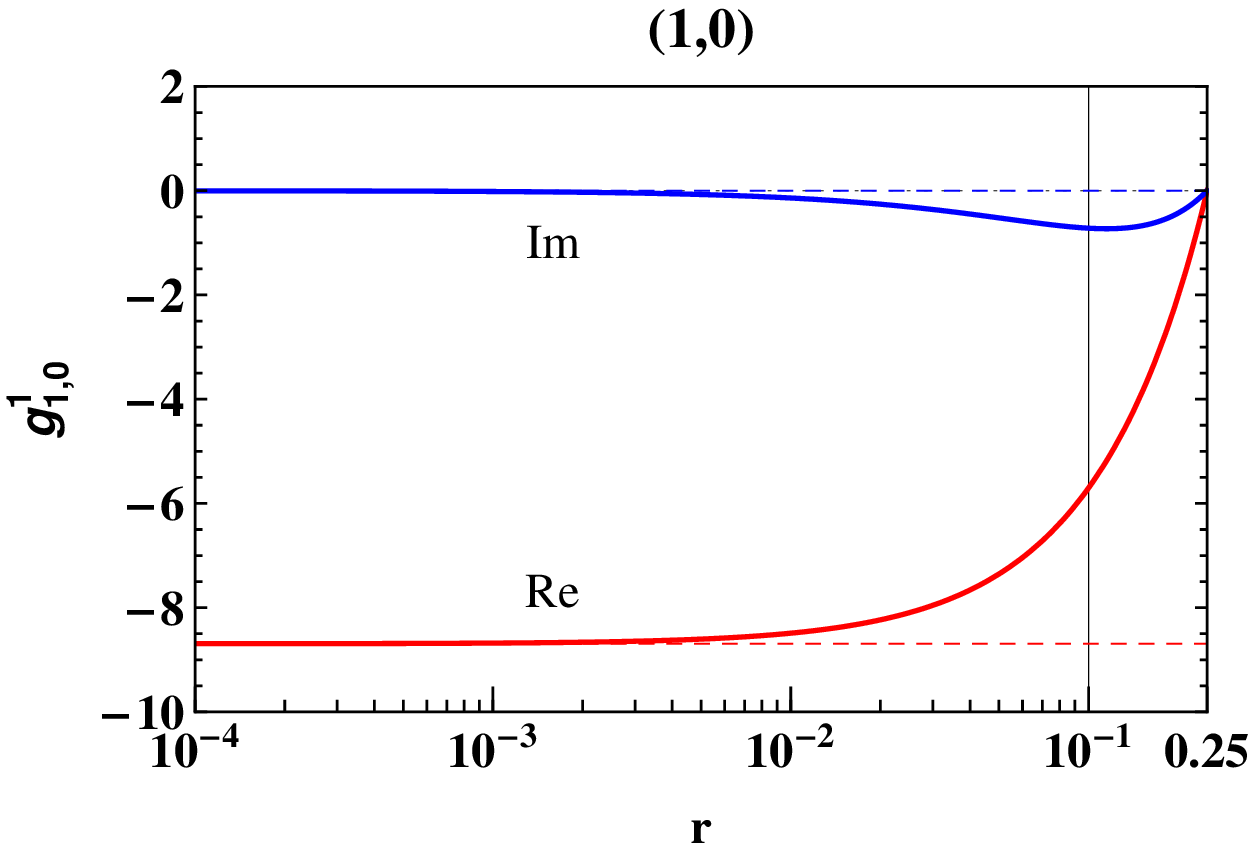}\ \ \
\includegraphics[width=0.3\textwidth]{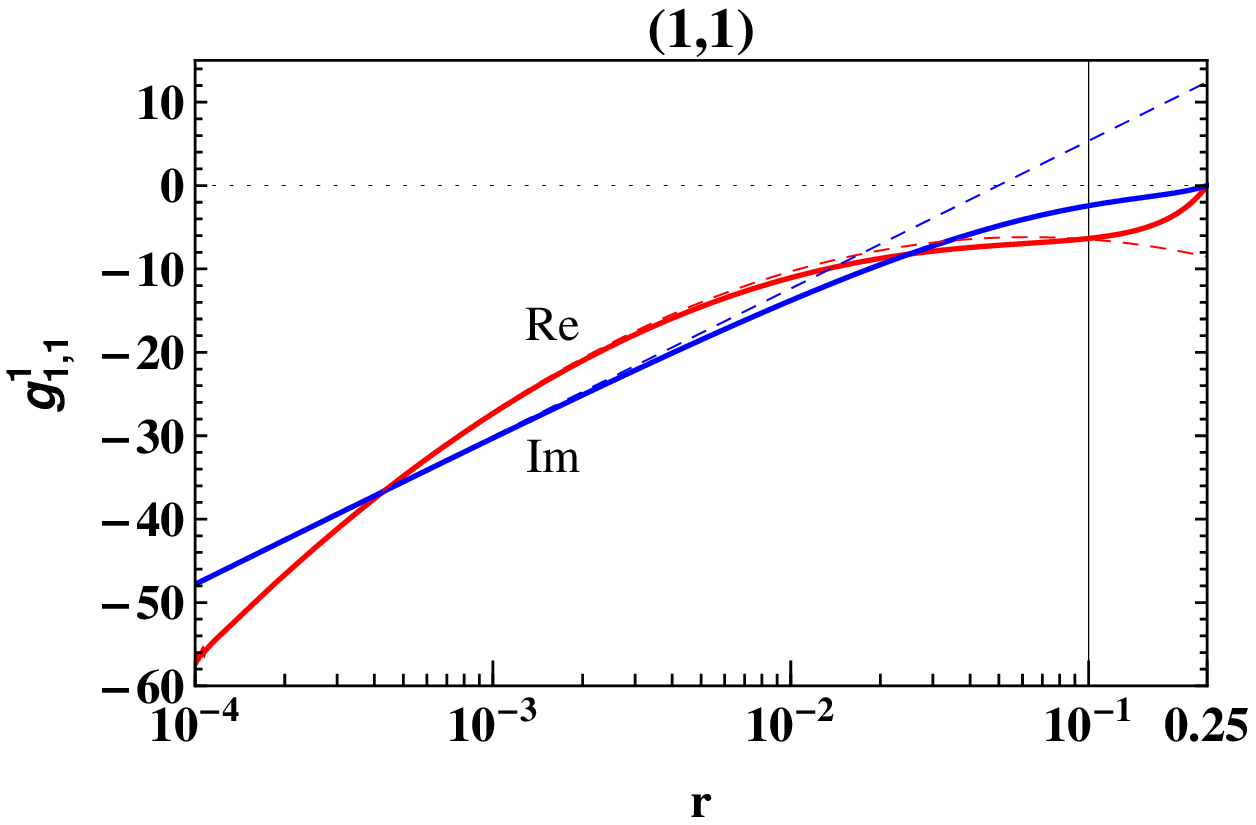}
\caption{Real and imaginary parts of
$g^1_{\lambda,\tilde{\lambda}}(r)$. The solid curves correspond to
the exact results, and the dashed curves represent the asymptotic
ones taken from (\ref{g:J:asym:expression}c) to
(\ref{g:J:asym:expression}e).
%--------------------
\label{plot:2g1gamma:g:1}}
%--------------------
\end{figure}

\begin{figure}[htb]
%\raggedright
\includegraphics[width=0.3\textwidth]{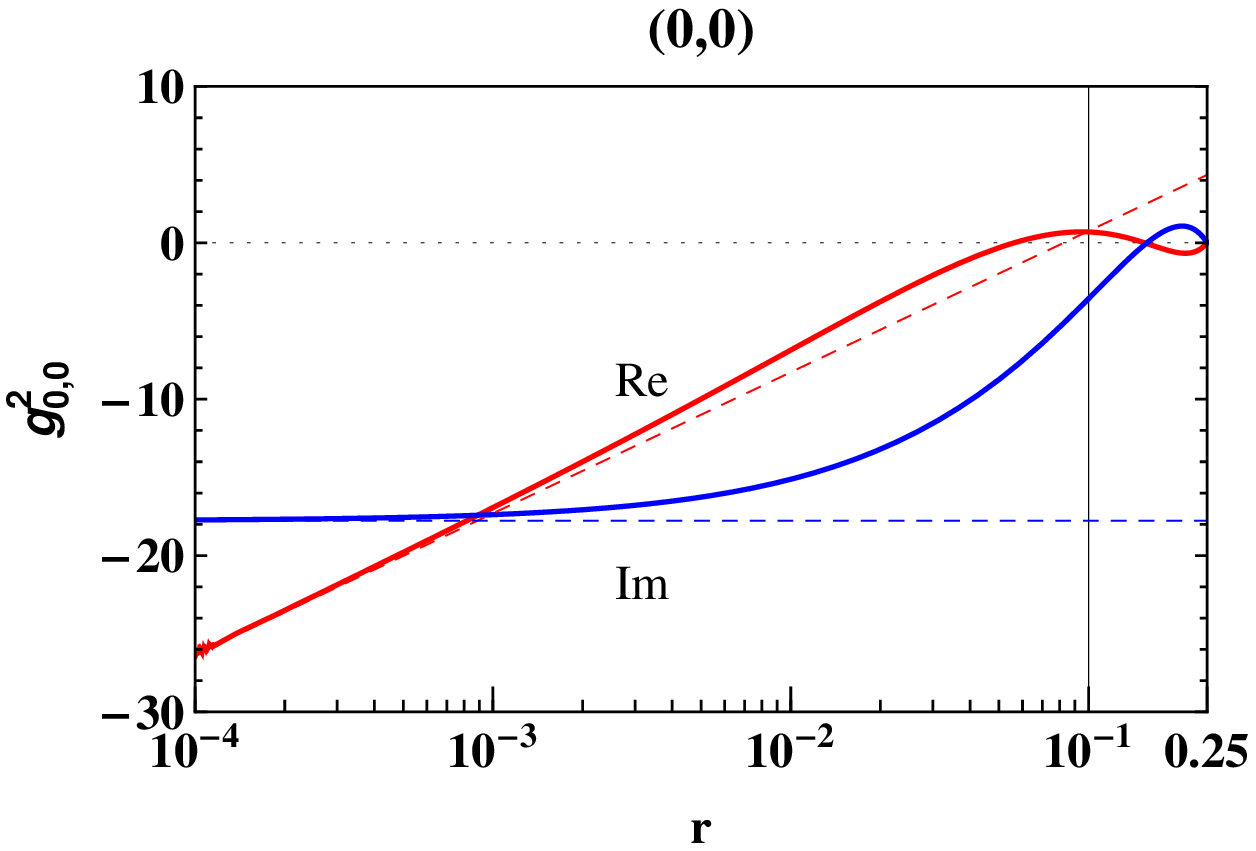}\ \ \
\includegraphics[width=0.3\textwidth]{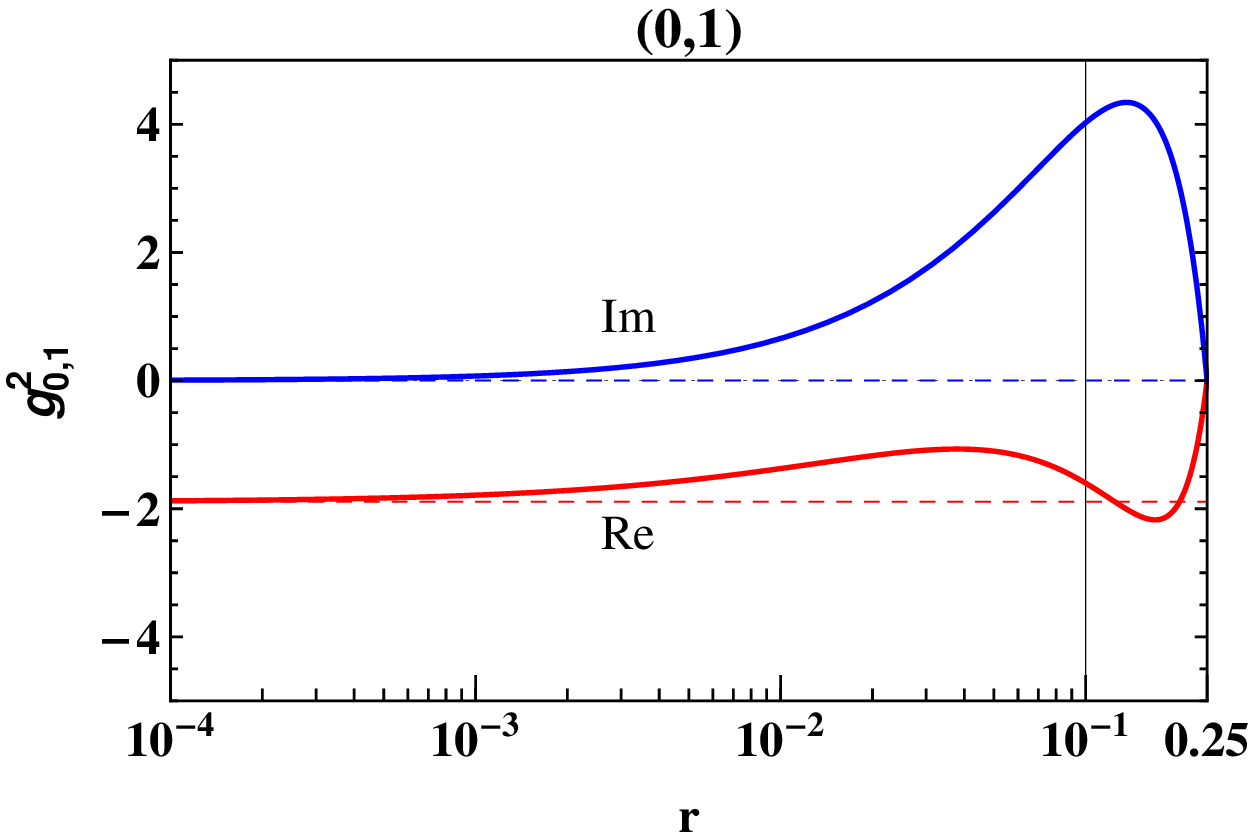}\ \ \
\includegraphics[width=0.3\textwidth]{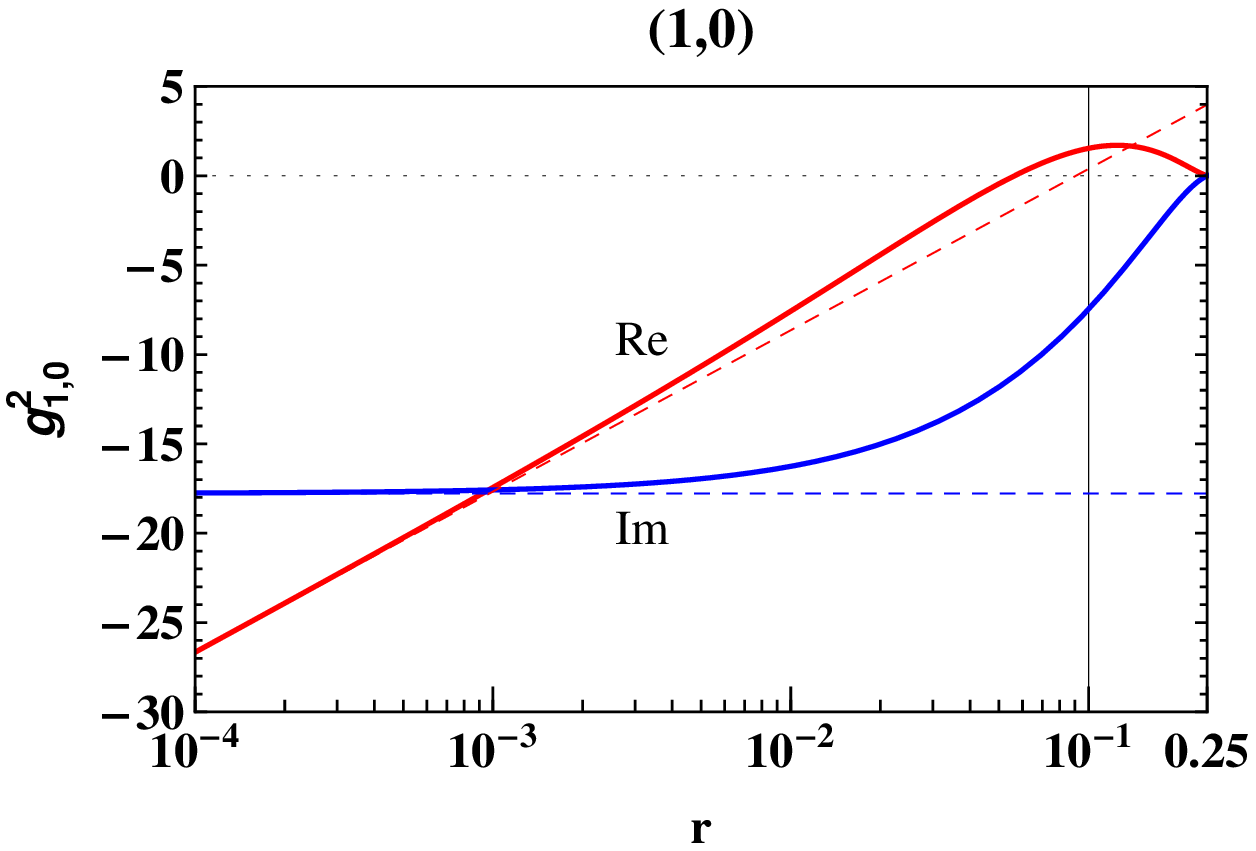}\ \ \
\includegraphics[width=0.3\textwidth]{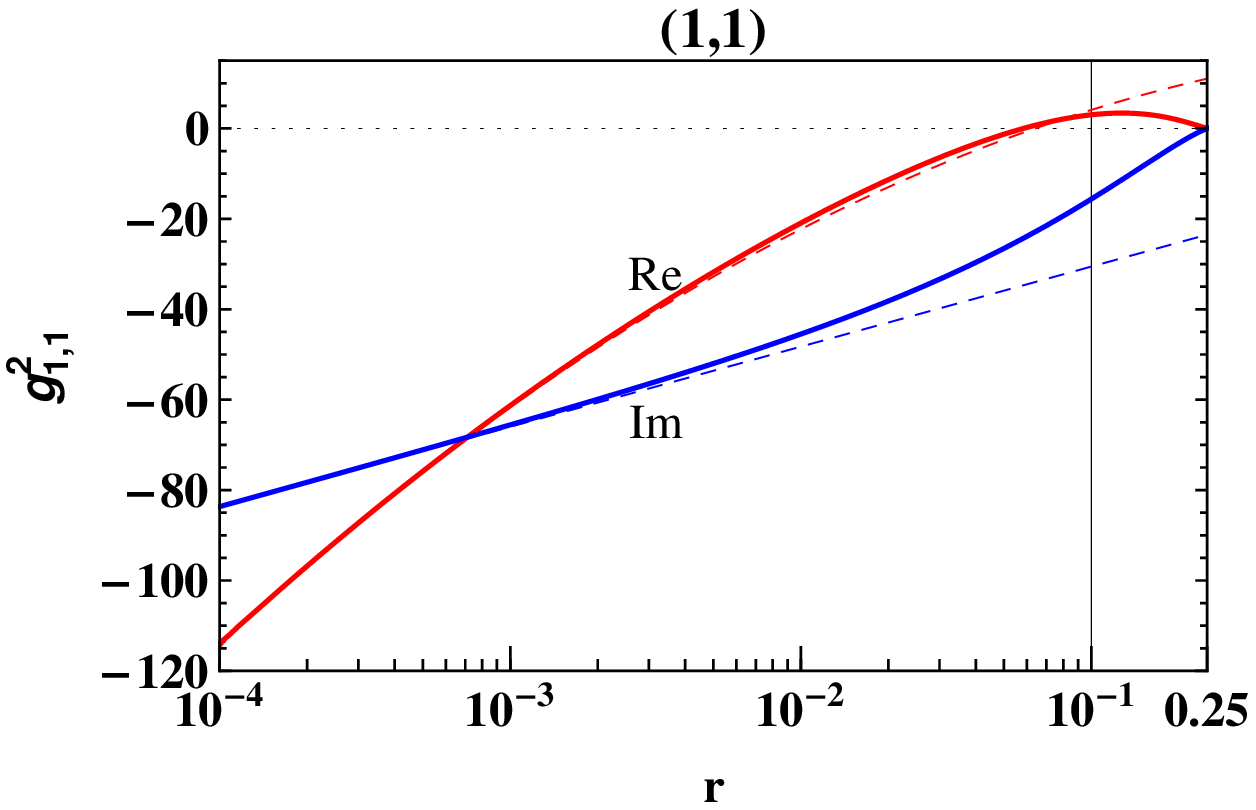}\ \ \
\includegraphics[width=0.3\textwidth]{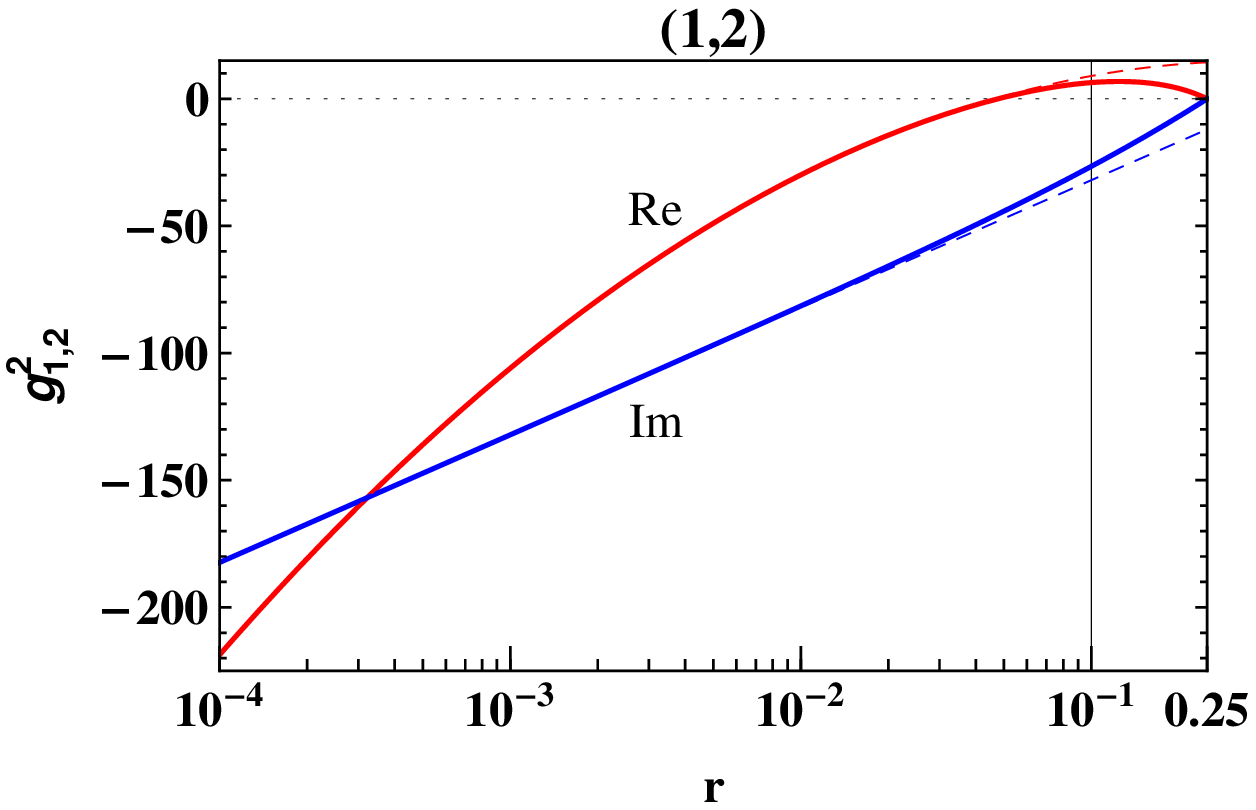}
\caption{Real and imaginary parts of
$g^2_{\lambda,\tilde{\lambda}}(r)$. The solid curves correspond to
the exact results, and the dashed curves represent the asymptotic
ones taken from (\ref{g:J:asym:expression}f) to
(\ref{g:J:asym:expression}j).
%--------------------
\label{plot:2g1gamma:g:2}}
%--------------------
\end{figure}

All the loop effects are encoded in the complex-valued,
dimensionless functions $ g^{J}_{\lambda,\tilde{\lambda}}(r)$. Their
full expressions are somewhat lengthy, and will not be reproduced
here. On the other hand, the profiles of these functions over a wide
range of $r$ are shown in Figs.~\ref{plot:2g1gamma:g:0},
\ref{plot:2g1gamma:g:1} and \ref{plot:2g1gamma:g:2}.

%--------------------
%--------------------

%--------------------
%--------------------

%--------------------
%--------------------

It is curious to know the asymptotic behavior of the reduced
helicity amplitudes in this decay channel. After some manipulations,
we find the asymptotic expressions of the functions
$g^J_{\lambda,\tilde{\lambda}}$ ($J=0,1,2$) to read
%-------------------
\begin{subequations}
%-------------------
\bqa
%-------------------
& & g^0_{0,0}(r) = 4 \ln 2 \ln r -{\pi^2\over 3} - 4\ln 2 + {53\over
9} +4 \pi i + {\cal O}(r\ln^2{r}),
%-------------------
\\
%-------------------
& & g^0_{1,0}(r) = 4\ln2 \ln r +{5\pi^2\over 12} -8\ln 2 +1 +4 \pi i
+ {\cal O}(r\ln^2{r}),
%-------------------
\\
%-------------------
& & g^1_{0,1}(r)= -4\sqrt{6}\left(\ln 2+1\right)+{\cal O}(r\ln{r}),
%-------------------
\\
%-------------------
& & g^1_{1,0}(r) = \sqrt{6}\left(2\ln 2-{\pi^2\over 2}\right) + {\cal O}(r\ln{r}),
%-------------------
\\
%-------------------
& & g^1_{1,1}(r) = -\sqrt{6}\left[\frac{1}{2}\ln^2 r +4\ln2\ln r
+{\pi^2 \over 12} +8\ln2 -\pi i \left(\ln r+3\right) \right] +{\cal
O}(r\ln^2{r}),
%-------------------
\\
%-------------------
& & g^2_{0,0}(r) = \sqrt{2} \left(4\ln{2}\ln{r} -{\pi^2\over 3}
+12\ln2 -{17\over 9} -4 \pi i \right)+ {\mathcal O}(r\ln^2r),
%-------------------
\\
%-------------------
& & g^2_{0,1}(r) = \sqrt{6}\left( 2 - 4\ln{2} \right) +{\cal O}(r\ln{r}),
%-------------------
\\
%-------------------
& & g^2_{1,0}(r) = \sqrt{2} \left( 4\ln 2\ln r +{5\pi^2\over 12}
+8\ln 2 -3 -4 \pi i \right) + {\cal O}(r\ln{r}),
%-------------------
\\
%-------------------
& & g^2_{1,1}(r)  =  \sqrt{6}\left[ -\frac{1}{2}\ln^2r -4\ln2\ln r
+4\ln r + {5\pi^2\over 12} -{20\over 3}\ln 2 +{23\over 3} + \pi i
\left(\ln r-\frac{5}{3}\right) \right]
%-------------------
\nn \\
%-------------------
& &  \qquad \quad +{\cal O}(r\ln^2{r}),
%-------------------
\\
%-------------------
& & g^2_{1,2}(r) = \sqrt{3}\left[ -2\ln^2 r -4\ln r +{5\pi^2\over 6}
+{16\over 3}\ln 2 -{16\over 3} +2 \pi i \left(2\ln{r}+{5\over
3}\right) \right]
%-------------------
\nn \\
%-------------------
&& \qquad \quad +{\cal O}(r\ln{r}).
%-------------------
\eqa
%-------------------
\label{g:J:asym:expression}
%-------------------
\end{subequations}
%-------------------

In Figs.~\ref{plot:2g1gamma:g:0}, \ref{plot:2g1gamma:g:1},
\ref{plot:2g1gamma:g:2}, we also juxtapose these asymptotic results
of $g^J_{\lambda,\tilde{\lambda}}(r)$ with the exact results. For
most helicity configurations, the asymptotic results seem not to
converge well with the exact ones for the phenomenologically
relevant point $r=m_c^2/m_b^2\approx 0.10$.

A quick survey on (\ref{g:J:asym:expression}) reveals that that the
scaling violation is again of the logarithmic form. More
interestingly, the same pattern of double logarithms still holds:
the occurrence of the double-logarithm $\ln^2 r$ is always
affiliated with the helicity-suppressed decay channels.

We close this section by making a simple observation. As can be seen
from Fig.~\ref{plot:2g1gamma:g:1}, the imaginary parts of the
functions $g^1_{\lambda,\tilde{\lambda}}$ are generally nonzero,
though some of which vanish asymptotically. At first sight, this may
contradict Landau-Yang theorem because $\chi_{c1}\to gg$ should be
strictly forbidden. This implies that, though the apparent two-gluon
cuts in Fig.~\ref{feynman:diag:gammagg} do not contribute, the other
cuts that simultaneously pass through the bottom and charm quark
lines must yield nonvanishing contributions to the imaginary parts
for $\Upsilon \rightarrow J/\psi+\chi_{c1}$.

%%%%%%%%%%%%%%%%%%%%%%%%%%%%%%%%%
\section{phenomenology}
\label{phenomenology}
%%%%%%%%%%%%%%%%%%%%%%%%%%%%%%%%%

We are now in a position to make concrete predictions for the decay
rates of $\Upsilon\to J/\psi+\chi_{c0,1,2}$, by plugging
(\ref{channel:gamma:red:hel:ampl}), (\ref{channel:3g:red:hel:ampl})
and (\ref{channel:gammagg:red:hel:ampl}) into
(\ref{polar:decay:rate:reduced:hel:ampl}).

In the numerical analysis, we take the various bottomonia and
charmonia masses from the 2012 PDG
compilation~\cite{Beringer:1900zz}: $M_{\Upsilon(1S)}=9.460$ GeV,
$M_{\Upsilon(2S)}=10.023$ GeV, $M_{\Upsilon(3S)}=10.355$ GeV,
$M_{J/\psi}=3.097$ GeV, $M_{\chi_{c0}}=3.415$ GeV,
$M_{\chi_{c1}}=3.511$ GeV, $M_{\chi_{c2}}=3.556$ GeV. These inputs
are used to determine $|{\bf P}|$ according to
(\ref{3:momentum:Kallen:function}), which appears in the phase space
factor in (\ref{polar:decay:rate:reduced:hel:ampl}). To calculate
the squared matrix elements, particularly the functions
$f^J_{\lambda,\widetilde{\lambda}}(r)$ and
$g^J_{\lambda,\widetilde{\lambda}}(r)$, we instead adopt the
following values for the quark masses: $m_c=1.5\ \mathrm{GeV}$,
$m_b=4.7\ \mathrm{GeV}$, corresponding to $r=0.102$. Later we also
wish to predict the branching fractions for the decays
$\Upsilon(nS)\to J/\psi+\chi_{c0,1,2}$ ($n=1,2,3$). For this
purpose, we take the total decay width of various $\Upsilon$ states
from \cite{Beringer:1900zz}: $\Gamma[\Upsilon(1S)]=54.02$ keV,
$\Gamma[\Upsilon(2S)]=31.98$ keV, and $\Gamma[\Upsilon(3S)]=20.32$
keV, respectively.

For simplicity, we fix the electromagnetic fine structure constant
as $\alpha=1/137$. For the strong coupling constant, any value
evaluated with the renormalization scale ranging from $2m_b$ to
$2m_c$ seems to be acceptable. This scale ambiguity constitutes one
of the most important sources of theoretical uncertainty. Without
much prejudice, we simply choose a medium value $\alpha_s(m_b)=
0.22$.

As for the wave functions at the origin for various quarkonia, we
use the values obtained from the Buchm\"{u}ller-Tye potential
model~\cite{Eichten:1995}: $|R_{\Upsilon (1S)}(0)|^2=6.477\
\mathrm{GeV}^3$, $|R_{\Upsilon (2S)}(0)|^2=3.234\ \mathrm{GeV}^3$,
$|R_{\Upsilon (3S)}(0)|^2=2.474\ \mathrm{GeV}^3$,
$|R_{J/\Psi}(0)|^2=0.81\ \mathrm{GeV}^3$,
$|R'_{\chi_{c0,1,2}}(0)|^2=0.075\ \mathrm{GeV}^5$.

%-----------------------------------------------
\begin{table*}[ht]
%-----------------------------------------------
\caption{The values of the reduced helicity amplitudes $a_{\gamma}$,
$a_{3g}$ and $a_{\gamma gg}$ associated with each helicity channel
for $\Upsilon(1S)\to J/\psi+\chi_{c0,1,2}$. For simplicity, a factor
of $10^{-4}$ has been pulled out of each entry.}
%-----------------------------------------------
\begin{center}
%-----------------------------------------------
\begin{tabularx}{\textwidth}{C{0.12\textwidth}XC{0.15\textwidth}C{0.15\textwidth}C{0.15\textwidth}C{0.15\textwidth}C{0.15\textwidth}}
\hlinew{1pt}
    &   & (0,0)              & (1,0) &(0,1)   &  (1,1)  &  (1,2) \\
\hline \multirow{3}{*}{$J/\psi+\chi_{c0}$}
 & $a_{\gamma}$     & $-5.99$ & $-24.23$&    --    &    --   &     --   \\
 & $a_{3g}$         & $18.33e^{i0.54^{\circ}}$  & $82.87e^{-i3.19^{\circ}}$&    --    &    --   &     --   \\
 & $a_{\gamma gg}$  & $1.58e^{-i66.32^{\circ}}$  & $1.01e^{-i58.59^{\circ}}$&    --
 &    --   &     --   \\
\hline \multirow{3}{*}{ $J/\psi+\chi_{c1}$}
& $a_{\gamma}$     &   --   &$-0.35$&$9.61$&$9.27$&     --   \\
 &$a_{3g}$         &   --   &$78.04e^{i109.57^{\circ}}$ & $54.94e^{i112.26^{\circ}}$  & $135.67e^{i108.61^{\circ}}$ &     --   \\
  &$a_{\gamma gg}$         &   --   & $5.23e^{i7.29^{\circ}}$ & $1.30e^{i7.48^{\circ}}$
  & $6.20e^{i20.52^{\circ}}$ &     --   \\
\hline \multirow{3}{*}{ $J/\psi+\chi_{c2}$}
& $a_{\gamma}$     & $-2.98$&$-8.94$&$-3.68$&$-11.80$&$-12.78$\\
 & $a_{3g}$\ \        & $1.15e^{i157.72^{\circ}}$& $30.89e^{i51.71^{\circ}}$ & $10.94e^{-i110.67^{\circ}}
 $& $57.18e^{i49.00^{\circ}}$& $120.16e^{i58.27^{\circ}}$ \\
  & $a_{\gamma gg}$\ \        & $0.33e^{i101.36^{\circ}}$& $6.89e^{i102.07^{\circ}}$ & $0.41e^{-i68.19^{\circ}}
 $& $14.35e^{i101.39^{\circ}}$& $24.70e^{i103.90^{\circ}}$ \\
\hlinew{1pt}
%-----------------------------------------------
\end{tabularx}\label{numerical:3:red:hel:ampl}
%-----------------------------------------------
\end{center}
%-----------------------------------------------
\end{table*}
%-----------------------------------------------

In Table~\ref{numerical:3:red:hel:ampl}, we tabulate the values of
the reduced helicity amplitudes $a_{\gamma}$, $a_{3g}$ and
$a_{\gamma gg}$ associated with each helicity channel for
$\Upsilon(1S)\to J/\psi+\chi_{c0,1,2}$. One sees that the relative
phases among each amplitude vary channel by channel, and
particularly there seems no universal phase pattern between the
single-photon and three-gluon channel. Our finding contradicts the
universal relative phase conjecture made in \cite{Gerard:1999uf}.

One curious question is whether the relative strength between three
distinct decay mechanisms bears the roughly same pattern as that in
$\Upsilon$ inclusive decay, as given in (\ref{Br:inclusive:ratio}).
A very crude guess is that each helicity amplitude for this
exclusive decay process might be proportional to $\sqrt{{\cal
B}_{\rm incl}}$, so that one naively expects
%-----------------------------------------------
\bqa
%-----------------------------------------------
& & |a_\gamma| :|a_{3g}|: |a_{\gamma gg}|\approx 1:3.3:0.5,
%-----------------------------------------------
\label{naive:guess:3:hel:ampl}
%-----------------------------------------------
\eqa
%-----------------------------------------------
for each helicity configuration.

Inspecting Table~\ref{numerical:3:red:hel:ampl}, we find that for
many helicity configurations, the three distinct helicity amplitudes
do exhibit the similar hierarchy as given in
(\ref{naive:guess:3:hel:ampl}), {\it i.e.}, $|a_{3g}|>
|a_{\gamma}|>|a_{\gamma gg}|$, though the radiative decay amplitude
are often much more suppressed. However, there are also a few
notable exceptions, {\it e.g.}, for the polarized decays $\Upsilon
\to J/\psi(\pm 1)+\chi_{c1}(0)$~\footnote{The unnaturally small
single-photon amplitude in this channel is due to the accidental
suppression factor received by $c^1_{1,0}$, as can be seen in
(\ref{c:function:gamma*:chic1}).}, $\Upsilon \to J/\psi(\pm
1)+\chi_{c2}(\pm 1)$, $\Upsilon \to J/\psi(\pm 1)+\chi_{c2}(\pm 2)$,
the magnitude of the radiative decay amplitudes is comparable, or
even greater, than that of the single-photon amplitudes; for the
decay $\Upsilon \to J/\psi(0)+\chi_{c2}(0)$, the single-photon
amplitude is even greater in magnitude than the respective
three-gluon amplitude.

%-----------------------------------------------
\begin{table*}[tb]
%-----------------------------------------------
\caption{The polarized and the polarization-summed partial widths
and the corresponding branching fractions. $\Gamma^{[J]}_{\gamma}$,
$\Gamma^{[J]}_{3g}$ and $\Gamma^{[J]}_{\gamma gg}$ represent the
individual decay rates from the single-photon, three-gluon, and
one-photon-two-gluon channels affiliated with the process
$\Upsilon(1S)\to J/\psi+\chi_{cJ}$ ($J=0,1,2$), respectively, while
$\Gamma^{[J]}_{\rm tot}$ denotes the decay rate which incorporates
all three decay mechanisms according to
(\ref{polar:decay:rate:reduced:hel:ampl}). The two rightmost columns
also give the unpolarized decay rates and branching fractions by
summing over all possible helicity configurations. All the partial
decay widths are in units of eV.}
%-----------------------------------------------
\begin{center}
%-----------------------------------------------
\begin{tabularx}{\textwidth}{X|C{0.13\textwidth}C{0.13\textwidth}C{0.13
\textwidth}C{0.13\textwidth}C{0.13\textwidth}C{0.13\textwidth}C{0.1\textwidth}}
\hlinew{1pt}
    & (0,0) & (1,0) &(0,1) &  (1,1)  & (1,2) & Unpol & ${\cal B}$ \\
\hline
  $ \Gamma^{[0]}_{\gamma}$     & $4.2\times 10^{-3}$ & $7.0\times 10^{-3}$&    --
  &    --   &     --  & $1.8\times 10^{-2}$ & $3.3\times 10^{-7}$ \\
  $\Gamma^{[0]}_{3g}$        &  $3.9\times 10^{-2}$  & $8.2\times 10^{-2}$&    --
  &    --   &     --  &  $2.0\times 10^{-1}$ & $3.7\times 10^{-6}$ \\
  $\Gamma^{[0]}_{\gamma gg}$&  $2.9\times 10^{-4}$  & $1.2\times 10^{-3}$&    --
   &    --   &     --  &  $2.6\times 10^{-3}$ & $4.8\times 10^{-8}$ \\
  $\Gamma^{[0]}_{\rm tot}$&  $1.4\times 10^{-2}$  & $2.8\times 10^{-2}$&    --
  &    --   &     --  &  $7.0\times 10^{-2}$ & $1.3\times 10^{-6}$ \\
\hline
 $\Gamma^{[1]}_{\gamma}$    &   --   &$1.4\times 10^{-6}$&$1.1\times 10^{-3}$&
 $1.0\times 10^{-4}$&     --  &$2.4\times 10^{-3}$ & $4.4\times 10^{-8}$ \\
 $\Gamma^{[1]}_{3g}$         &   --   &$7.1\times 10^{-2}$ & $3.5\times 10^{-2}$ &
 $2.2\times 10^{-2}$ &     --  &$2.6\times 10^{-1}$ & $4.8\times 10^{-6}$ \\
 $\Gamma^{[1]}_{\gamma gg}$&   --   &$3.2\times 10^{-4}$ & $2.0\times 10^{-5}$ &
 $4.6\times 10^{-5}$ &     --  &$7.7\times 10^{-4}$ & $1.4\times 10^{-8}$ \\
 $\Gamma^{[1]}_{\rm tot}$&  $-$  & $6.2\times 10^{-2}$&    $4.7\times 10^{-2}$    &
 $2.3\times 10^{-2}$   &     --  &  $2.6\times 10^{-1}$ & $4.9\times 10^{-6}$ \\
\hline
  $ \Gamma^{[2]}_{\gamma}$ &$1.0\times 10^{-3}$&$9.3\times 10^{-4}$&$1.6\times 10^{-4}$
   & $1.7\times 10^{-4}$&$2.0\times 10^{-5}$&$3.6\times 10^{-3}$ & $6.7\times 10^{-8}$ \\
  $ \Gamma^{[2]}_{3g}$ &$1.5\times 10^{-4}$&$1.1\times 10^{-2}$ & $1.4\times 10^{-3}$
  & $3.9\times 10^{-3}$&$1.7\times 10^{-3}$
  &$3.6\times 10^{-2}$ & $6.7\times 10^{-7}$ \\
    $\Gamma^{[2]}_{\gamma gg}$&$1.2\times 10^{-5}$&$5.5\times 10^{-4}$
    &$2.0\times 10^{-6}$&$2.4\times 10^{-4}$&$7.4\times 10^{-5}$
  &$1.8\times 10^{-3}$ & $3.3\times 10^{-8}$ \\
  $\Gamma^{[2]}_{\rm tot}$&  $5.3\times 10^{-4}$  & $2.6\times 10^{-3}$&
  $5.5\times 10^{-4}$ & $1.1\times 10^{-3}$   & $8.3\times 10^{-4}$  &
  $1.1\times 10^{-2}$ & $2.0\times 10^{-7}$ \\
\hlinew{1pt}
%-----------------------------------------------
\end{tabularx}\label{num:pol:decay:partial:widths}
%-----------------------------------------------
\end{center}
%-----------------------------------------------
\end{table*}
%-----------------------------------------------

In Table~\ref{num:pol:decay:partial:widths}, we also list the
polarized decay widths for $\Upsilon(1S)\to
J/\psi(\lambda)+\chi_{c0,1,2}(\tilde{\lambda})$ for each independent
helicity configurations $(\lambda,\tilde{\lambda})$, together with
the polarization-summed results. To readily visualize the
interference effect among three distinct decay mechanisms, we also
tabulate the individual decay rates from the single-photon,
three-gluon, and one-photon-two-gluon channels, respectively, as
well as the full decay rate given by
(\ref{polar:decay:rate:reduced:hel:ampl}).
Table~\ref{num:pol:decay:partial:widths} reveals that $a_\gamma$ and
$a_{3g}$ in $\Upsilon(1S)\to J/\psi+\chi_{c0,2}$ are subject to {\it
destructive} interference.

The hierarchy among the polarized decay widths in different helicity
configurations, from either an individual decay mechanism or the
complete contributions, seems hardly to obey the HSR as indicated in
(\ref{helicity:selection:rule}). The pattern is abnormal for
$\Upsilon\to J/\psi+\chi_{c0}$, where the helicity-suppressed $(\pm
1,0)$ channel even possesses a bigger decay rate than the
helicity-favored $(0,0)$ channel; for $\Upsilon\to
J/\psi+\chi_{c2}$, the largest polarized decay rates are associated
with the helicity-suppressed configurations $(\pm 1,0)$ and $(\pm
1,\pm 1)$, whereas the smallest polarized decay rate is associated
with the HSR-favored $(0,0)$ state~\footnote{Note this situation is
quite different from the continuum production process $e^+e^-\to
J/\psi(\lambda)+\chi_{c2}(\tilde{\lambda})$, where the $(0,0)$ and
$(\pm 1,0)$ channels make the dominant contributions to the
unpolarized production cross section~\cite{Dong:2011fb}.}. These
symptoms can presumably be attributed to the fact that the mass
ratio $m_c/m_b\approx 1/3$ might not be small enough to warrant the
asymptotic counting rule.

%-----------------------------------------------
\begin{table*}[tb]
%-----------------------------------------------
\caption{The unpolarized partial decay widths and the corresponding
branching fractions that incorporate all three distinct decay
mechanisms. The superscript $[J]$ characterizes the corresponding
decay process $\Upsilon(1S,2S,3S)\to J/\psi+\chi_{cJ}$ ($J=0,1,2$).}
%-----------------------------------------------
\begin{center}
%-----------------------------------------------
\begin{tabularx}{\textwidth}{XC{0.13\textwidth}C{0.14\textwidth}C
{0.14\textwidth}C{0.14\textwidth}C{0.14\textwidth}C{0.14\textwidth}}
\hlinew{1pt}
 & $\Gamma^{[0]}$ (eV) & ${\mathcal B}^{[0]}$ & $\Gamma^{[1]}$ (eV) &
 ${\mathcal B}^{[1]}$
 &$\Gamma^{[2]}$(eV) & ${\mathcal B}^{[2]}$ \\ \hline
 $\Upsilon$(1S)&$0.070 ^{}  $&$1.3\times 10^{-6}$&$0.26$&$4.9\times 10^{-6}$&$0.011$&$2.0\times 10^{-7}$\\
 \hline
  $\Upsilon$(2S)&$0.035  $&$1.1\times 10^{-6}$&$0.13$&$4.1\times 10^{-6}$&$0.0054$&$1.7\times 10^{-7}$\\
  \hline
   $\Upsilon$(3S)& $0.026 $& $8.6\times 10^{-7}$ &$0.099$ & $3.3\times 10^{-6}$&$0.0041$ & $1.3\times 10^{-7}$\\
\hlinew{1pt}
%-----------------------------------------------
\end{tabularx}\label{partial:widths:Upsilon:nS}
%-----------------------------------------------
\end{center}
%-----------------------------------------------
\end{table*}
%-----------------------------------------------

Finally in Table~\ref{partial:widths:Upsilon:nS}, we tabulate our
predictions of the partial decay widths, which are calculated
according to (\ref{unpol:decay:rate:Jpsi:chicJ}), together with the
corresponding branching fractions, for the various processes
$\Upsilon(nS)\to J/\psi+\chi_{c0,1,2}$ ($n=1,2,3$). All the three
decay mechanisms are incorporated. We observe that the decay
branching fractions satisfy the ordering~\footnote{It is interesting
to compare the relative importance of the various exclusive
production channels $J/\psi+\chi_{cJ}$ ($J=0,1,2$) in $\Upsilon$
decay and continuum production. For $e^+e^-\to
J/\psi+\chi_{c0,1,2}$, one finds that the production rate for
$J/\psi+\chi_{c0}$ is about one order of magnitude greater than
those for $J/\psi+\chi_{c1,2}$~\cite{Wang:2011qg,Dong:2011fb}.}
${\cal B}^{[1]}>{\cal B}^{[0]} \gg {\cal B}^{[2]}$, with the first
two reaching the order of $10^{-6}$. We note that all these
predicted branching ratios are compatible with the various
experimental bounds on $J/\psi$ or $\chi_{cJ}$ inclusive production
rates in $\Upsilon(1S,2S)$ decays, as given in
(\ref{inclusive:Psi:production:ups}).

As a simple consequence of the LO NRQCD prediction, there should
exist a 77\% rule in the hadronic decay of the $\Upsilon$ system, in
analogy with the famous 12\% rule in the $\psi$ system:
%-------------
\bqa
%-------------
 & & {{\cal B}[\Upsilon(2S) \to {\rm hadrons}] \over {\cal B}[\Upsilon(1S) \to {\rm hadrons}]}
  = {{\cal B}[\Upsilon(2S) \to e^+e^-] \over {\cal B}[\Upsilon(1S) \to e^+e^-]
 }
 = 0.77\pm 0.07.
%-------------
\label{80:percent:rule}
%-------------
\eqa
%-------------
Not surprisingly, the ratios of ${\cal B}[\Upsilon(2S)\to
J/\psi+\chi_{cJ}]$ to ${\cal B}[\Upsilon(1S)\to J/\psi+\chi_{cJ}]$
in Table~\ref{partial:widths:Upsilon:nS} are indeed compatible with
this rule.

Thus far, the \textsc{Belle} experiment has collected about $102$
million $\Upsilon(1S)$ samples and $158$ million $\Upsilon(2S)$
samples. According to Table~\ref{partial:widths:Upsilon:nS}, the
\textsc{Belle} experiment is expected to have produced about 130 and
170 $\Upsilon(1S,2S)\to J/\psi+\chi_{c0}$ events, 500 and 650
$\Upsilon(1S,2S)\to J/\psi+\chi_{c1}$ events, 20 and 30
$\Upsilon(1S,2S)\to J/\psi+\chi_{c2}$ events, respectively.

Experimentally, there are two possible methods to detect the $
J/\psi+\chi_{c0,1,2}$ signals. The first is to reconstruct both the
$J/\psi$ and $\chi_{c0,1,2}$ events. The clean and copious decay
modes of $\chi_{c1,2}$ are the $E1$ radiative transitions
$\chi_{c1,2}\to J/\psi+\gamma$, with the branching fractions of
$34.4$\% and $19.5$\%, respectively~\cite{Beringer:1900zz}. The
$J/\psi$ meson can be most cleanly tagged through the leptonic
decays into $e^+e^-$ and $\mu^+\mu^-$, with combined branching
ratios about 12\%~\cite{Beringer:1900zz}. Although this method has
the advantage of bearing very low background level, taking into
account the reconstruction efficiencies for two $J/\psi$ and one
photon, one may end up with too few signal events to be practically
useful.

The second method is to only reconstruct one $J/\psi\to l^+l^-$
event, then fit the recoil mass spectrum against the $J/\psi$ to
estimate the number of $\chi_{c0,1,2}$ peak events. This method will
not depend on the concrete decay modes of $\chi_{c0,1,2}$. For low
statistics of signal events like in our case, this method is much
more superior to the preceding one. As a matter of fact, this method
has already been used by the \textsc{Belle} collaboration to impose
the upper bound for the exclusive bottomonium decays
$\chi_{b0,1,2}\to J/\psi J/\psi, J/\psi \psi'$~\cite{Shen:2012ei}.

In fitting the recoil mass spectrum of $J/\psi$, the net detection
efficiency for $\Upsilon(1S,2S)\to J/\psi+ \chi_{c0,1,2}$ is
estimated to be around 4\% (similar for all $\chi_{c0,1,2}$), with
the reconstruction efficiency for $J/\psi\to l^+l^-$
included~\cite{Shen:2012:private}. Therefore, the numbers of the
observed $\Upsilon(1S,2S)\to J/\psi+ \chi_{c1}$ events are expected
to be $500\times 4\%=20$, and $650\times 4\%=26$, respectively.
Since only one $J/\psi$ is reconstructed, the background level in
real data may not be very low. With only 20 reconstructed signal
events, it seems quite challenging for the signal significance to
reach the $5 \sigma$ level, and a larger data pool is needed in
order to draw a definite conclusion. In the prospective Super $B$
factory, with a luminosity 50 times greater than the current $B$
factory, it seems very promising that the decays
$\Upsilon(1S,2S,3S)\to J/\psi+\chi_{c0,1}$ will be eventually
observed.

\section{Summary}
\label{summary}

In this paper, we carry out a comprehensive investigation on the
exclusive $J/\psi+\chi_{c0,1,2}$ production in $\Upsilon$ decay in
the NRQCD factorization framework. We have explicitly considered
three distinct decay mechanisms, {\it i.e.}, the strong,
electromagnetic and radiative decay channels. Although there has not
yet appeared a rigorous proof on the validity of NRQCD factorization
approach to these types of double-charmonium production processes,
the explicit verification for the cancelation of IR divergences in
our calculation is rather supportive of the positive answer.
Moreover, our explicit calculation further supports the previous
claim that the double logarithms appearing in the one-loop NRQCD
short-distance coefficients are always affiliated with the
helicity-suppressed channels~\cite{Jia:2010fw,Dong:2011fb}.

The branching fractions for the various polarized and unpolarized
decay channels $\Upsilon(nS)\to J/\psi+\chi_{c0,1,2}$ ($n=1,2,3$)
are predicted by incorporating all three distinct decay channels at
the lowest order. In our case, the relative phase among these decay
channels arise from the short-distance loop effect, which can
actually be calculated in perturbation theory. There appears no
universal interference pattern, but the three-gluon and the
single-photon amplitudes often tend to be destructive. We find that
$\Upsilon(nS)$ decays into $J/\psi+\chi_{c1}$ have the largest
branching fraction, about a few times $10^{-6}$; and the decays
$\Upsilon(nS)\to J/\psi+\chi_{c2}$ have the smallest decay branching
ratio, only of order $10^{-7}$. The current statistics at
\textsc{Belle} is on the margin of observing these decay channels.
If the prospective high-luminosity $e^+e^-$ facilities such as the
Super B experiment can dedicate more machine time on the first three
$\Upsilon$ resonances, it should be an ideal place to discover their
exclusive decay modes into $J/\psi+\chi_{c0,1}$.

%--------------------------------------------------------------------
\begin{acknowledgments}
%--------------------------------------------------------------------
 We are grateful to Cheng-Ping Shen and Chang-Zheng Yuan for the
 nice explanations of the experimental issues about detecting the
 $J/\psi+\chi_{cJ}$ signals in
 \textsc{Belle} experiment.
%--------------------------------------------------------------------
This research was supported in part by the National Natural Science
Foundation of China under Grant Nos.~10875130, 10875156, 10935012,
11125525, DFG and NSFC (CRC 110), and by the Ministry of Science and
Technology of China under Contract No. 2009CB825200.
%--------------------------------------------------------------------
\end{acknowledgments}
%--------------------------------------------------------------------

\appendix

%--------------------------------------------------------------------
\section{Various helicity projectors for $\Upsilon\to J/\psi+\chi_{c0,1,2}$
\label{helicity:projectors}%
}

During this work, we have utilized the various helicity projectors
to expedite projecting out the corresponding helicity amplitudes
associated with $\Upsilon\to J/\psi+\chi_{c0,1,2}$. This helicity
projection technique, which has already been applied in our previous
work on the ${\mathcal O}(\alpha_s)$ correction to the process
$e^+e^-\to J/\psi+\chi_{c0,1,2}$~\cite{Dong:2011fb}, can
significantly reduce the amount of labors required for the loop
diagram calculations. In this Appendix, we collect the explicit
formulas for the 10 helicity projectors used in this work.

Firstly, it is convenient to introduce the {\it transverse} metric
tensor:
%--------------------------------
\bqa
%--------------------------------
g_{\perp\:\mu\nu} &\equiv& g_{\mu\nu}+{P_\mu P_\nu \over |{\bf
P}|^2}- {Q\cdot P\over M^2_{\Upsilon}|{\bf P}|^2} (P_\mu Q_\nu+Q_\mu
P_\nu)+{M_{J/\psi}^2\over M^2_{\Upsilon}} {Q_\mu Q_\nu \over |{\bf
P}|^2}
%--------------------------------
\nn \\
%--------------------------------
&=&  g_{\mu\nu}+{\widetilde{P}_\mu \widetilde{P}_\nu \over |{\bf
P}|^2}- {Q\cdot \widetilde{P}\over M^2_{\Upsilon}|{\bf P}|^2}
(\widetilde{P}_\mu Q_\nu+Q_\mu
\widetilde{P}_\nu)+{M_{\chi_{cJ}}^2\over M^2_{\Upsilon}} {Q_\mu
Q_\nu \over |{\bf P}|^2},
%--------------------------------
\label{g_perp:def}
%--------------------------------
\eqa
%--------------------------------
where $Q$, $P$ and $\widetilde{P}$ stand for the four-momenta of
$\Upsilon$, $\jpsi$ and $\chi_{cJ}$, respectively. This symmetric
tensor satisfies the transversity condition $g_{\perp\:\mu\nu} P^\mu
=g_{\perp\:\mu\nu} \widetilde{P}^\mu=0$.
%--------------------------------
It further has the properties $g_{\perp\:\mu}^{\;\;\mu}=2$,
$g_{\perp\:\mu\alpha}\, g_{\perp}^{\alpha\nu}=
g_{\perp\:\mu\alpha}\, g^{\alpha\nu}= g_{\perp\:\mu}^{\;\;\;\nu}$.

The decay amplitude ${\mathcal A}^0$ for the process
$\Upsilon\rightarrow J/\psi+\chi_{c0}$ can be parameterized as
${\mathcal A}^{0}={\mathcal A}_{\mu\nu}^0\epsilon^\mu_\Upsilon
\epsilon^{*\nu}_{J/\psi}(\lambda)$, where $\epsilon_\Upsilon$ and
$\epsilon_{J/\psi}$ denote the polarization vectors of $\Upsilon$
and $J/\psi$, respectively. The helicities of $J/\psi$ and
$\chi_{c0}$ are labeled by $\lambda$ and $\tilde{\lambda}$
(trivially $\tilde{\lambda}=0$). There are only two independent
helicity amplitudes for this process, which can be deduced by acting
the corresponding helicity projectors on the amputated amplitude:
${\mathcal A}_{\lambda, \tilde{\lambda}}^{0} = {\mathbb P}_{\lambda,
\tilde{\lambda}}^{\mu\nu} \, {\mathcal A}_{\mu\nu}^{0}$, up to an
immaterial phase. The two helicity projection tensors for
$\Upsilon\rightarrow \jpsi+\chi_{c0}$ are
%--------------------------------------------------------------
\begin{subequations}
%--------------------------------------------------------------
\bqa
%--------------------------------------------------------------
{\mathbb P}_{0,0}^{\mu\nu} &=&   {1\over |{\bf P}|^2} \bigg(P_\mu -
{Q\cdot P\over M^2_{\Upsilon}}Q_\mu
 \bigg) \bigg({Q\cdot P\over M_{J/\psi}M_{\Upsilon}}P_\nu-
 {M_{J/\psi}\over M_{\Upsilon}} Q_\nu \bigg),
%--------------------------------------------------------------
\\
%--------------------------------------------------------------
{\mathbb P}_{1,0}^{\mu\nu} &=& -{1\over 2}\,g_{\perp\:\mu\nu},
%--------------------------------------------------------------
\eqa
%--------------------------------------------------------------
\end{subequations}
%--------------------------------------------------------------
where $g_{\perp\:\mu\nu}$ is defined in (\ref{g_perp:def}). These
two projectors are normalized as ${\mathbb P}_{0,0;\mu\nu}\,
{\mathbb P}_{0,0}^{\mu\nu}=1$, ${\mathbb P}_{1,0;\mu\nu}\,{\mathbb
P}_{1,0}^{\mu\nu}=\frac{1}{2}$, and orthogonal to each other:
${\mathbb P}_{0,0;\mu\nu}\,{\mathbb P}_{1,0}^{\mu\nu}=0$.

The decay amplitude ${\mathcal A}^{1}$ for the decay process
$\Upsilon\rightarrow J/\psi+\chi_{c1}$ can be expressed as
${\mathcal A}^{1}={\mathcal
A}_{\mu\nu\alpha}^{1}\epsilon^\mu_\Upsilon
\epsilon^{*\nu}_{J/\psi}(\lambda)
\epsilon^{*\alpha}_{\chi_{c1}}(\tilde{\lambda})$, where
$\tilde{\lambda}$ and $\epsilon_{\chi_{c1}}$ denote the helicity of
the $\chi_{c1}$ state and the corresponding polarization vector.
There are three independent helicity amplitudes for this process,
which can be deduced by acting the corresponding helicity projectors
on the amputated amplitude: ${\mathcal A}_{\lambda,
\tilde{\lambda}}^{1} = {\mathbb P}_{\lambda,
\tilde{\lambda}}^{\mu\nu\alpha} \,{\mathcal A}_{\mu\nu\alpha}^{1}$.
The three helicity projection tensors for $\Upsilon\rightarrow
J/\psi+\chi_{c1}$ read:
%--------------------------------------------------------------
\begin{subequations}
%--------------------------------------------------------------
\bqa
%--------------------------------------------------------------
{\mathbb P}_{1,0}^{\mu\nu\alpha} &=&  {i\over 2 M_{\Upsilon}  |{\bf
P}|^2}\, \epsilon_{\mu\nu\rho\sigma} Q^\rho \widetilde{P}^\sigma
\left({Q\cdot \widetilde{P}\over M_{\chi_{c1}}M_{\Upsilon}} {\tilde
P}_\alpha -
 {M_{\chi_{c1}}\over M_{\Upsilon}} Q_\alpha  \right),
%--------------------------------------------------------------
\\
%--------------------------------------------------------------
{\mathbb P}_{0,1}^{\mu\nu\alpha} &=& - {i\over 2 M_{\Upsilon}  |{\bf
P}|^2}\, \epsilon_{\mu\alpha\rho\sigma} Q^\rho P^\sigma \left(
{Q\cdot P\over M_{J/\psi}M_{\Upsilon}}  P_\nu -
 {M_{J/\psi}\over M_{\Upsilon}} Q_\nu  \right),\
%--------------------------------------------------------------
\\
%--------------------------------------------------------------
{\mathbb P}_{1,1}^{\mu\nu\alpha} &=& {i\over 2 M_{\Upsilon}  |{\bf
P}|^2}\, \epsilon_{\nu\alpha\rho\sigma} Q^\rho P^\sigma \left(P_\mu
- { Q\cdot P\over M^2_{\Upsilon}} Q_\mu  \right),
%--------------------------------------------------------------
\eqa
%--------------------------------------------------------------
\end{subequations}
%--------------------------------------------------------------
which are subject to the normalization conditions ${\mathbb
P}_{i;\mu\nu\alpha}\,{\mathbb P}_{j}^{\mu\nu\alpha}={1\over
2}\delta_{ij}$, with $i$, $j$ signifying one of the three helicity
configurations.

Similarly, for the decay process $\Upsilon\rightarrow
J/\psi+\chi_{c2}$, we can identify the amputated amplitude through
${\mathcal A}^{2}={\mathcal
A}_{\mu\nu\alpha\beta}^{2}\epsilon^\mu_\Upsilon
\epsilon^{*\nu}_{J/\psi}(\lambda)
e^{*\alpha\beta}_{\chi_{c2}}(\tilde{\lambda})$, where
$\tilde{\lambda}$ and $e_{\chi_{c2}}$  represent the helicity of the
$\chi_{c2}$ state and the corresponding polarization tensor. There
are in total five independent helicity amplitudes, which can be
obtained by acting the corresponding helicity projectors upon the
amputated amplitude: ${\mathcal A}_{\lambda,\tilde{\lambda}}^{2} =
{\mathbb P}_{\lambda,\tilde{\lambda}}^{\mu\nu\alpha\beta}
\,{\mathcal A}_{\mu\nu\alpha\beta}^{2}$. We construct the five
helicity projection tensors for $\Upsilon\rightarrow
\jpsi+\chi_{c2}$ as
%--------------------------------------------------------------
\begin{subequations}
%--------------------------------------------------------------
\bqa
%--------------------------------------------------------------
{\mathbb P}_{0,0}^{\mu\nu\alpha\beta} &=& {1 \over \sqrt{6}|{\bf
P}|^2 } \bigg({Q\cdot P\over M_{J/\psi}M_{\Upsilon}}P_\nu-
 {M_{J/\psi}\over M_{\Upsilon}} Q_\nu
 \bigg)
\bigg(P_\mu - {Q\cdot P\over M^2_{\Upsilon}}Q_\mu \bigg) \Bigg[
g_{\perp\:\alpha\beta} + {2\over |{\bf P}|^2 }
%--------------------------------------------------------------
\nn\\
%--------------------------------------------------------------
&& \times \bigg({Q\cdot \widetilde{P}\over
M_{\chi_{c2}}M_{\Upsilon}} {\widetilde P}_\alpha-
 {M_{\chi_{c2}}\over M_{\Upsilon}} Q_\alpha \bigg) \bigg({Q\cdot \widetilde{P}\over
M_{\chi_{c2}}M_{\Upsilon}} {\widetilde P}_\beta -
 {M_{\chi_{c2}}\over M_{\Upsilon}} Q_\beta  \bigg) \Bigg],
%--------------------------------------------------------------
\\
%--------------------------------------------------------------
{\mathbb P}_{1,0}^{\mu\nu\alpha\beta} &=& - {1 \over 2\sqrt{6}}
\,g_{\perp\:\mu\nu}  \Bigg[ g_{\perp\:\alpha\beta}+ {2\over |{\bf
P}|^2 }\bigg({Q\cdot \tilde{P}\over M_{\chi_{c2}}M_{\Upsilon}}
{\widetilde P}_\alpha -
 {M_{\chi_{c2}}\over M_{\Upsilon}} Q_\alpha \bigg) \bigg({Q\cdot \tilde{P}\over
M_{\chi_{c2}}M_{\Upsilon}} {\widetilde P}_\beta -
 {M_{\chi_{c2}}\over M_{\Upsilon}} Q_\beta \bigg) \Bigg],\nn
%--------------------------------------------------------------
\\ \\
%--------------------------------------------------------------
{\mathbb P}_{0,1}^{\mu\nu\alpha\beta} &=&  {1\over 2\sqrt{2}|{\bf
P}|^2 } \bigg( {Q\cdot P\over M_{J/\psi}M_{\Upsilon}}P_\nu-
 {M_{J/\psi}\over M_{\Upsilon}} Q_\nu
\bigg) \Bigg[g_{\perp\:\mu\alpha} \bigg({Q\cdot \widetilde{P}\over
M_{\chi_{c2}}M_{\Upsilon}} {\widetilde P}_\beta -
 {M_{\chi_{c2}}\over M_{\Upsilon}} Q_\beta \bigg)
%--------------------------------------------------------------
\nn \\
%--------------------------------------------------------------
& & +(\alpha \leftrightarrow \beta) \Bigg],
%--------------------------------------------------------------
\\
%--------------------------------------------------------------
{\mathbb P}_{1,1}^{\mu\nu\alpha\beta} &=& -{1\over 2\sqrt{2} |{\bf
P}|^2} \bigg(P_\mu - {Q\cdot P\over M^2_{\Upsilon}}Q_\mu \bigg)
\Bigg[g_{\perp\:\nu\alpha} \bigg( {Q\cdot \widetilde{P}\over
M_{\chi_{c2}}M_{\Upsilon}} {\widetilde P}_\beta -
 {M_{\chi_{c2}}\over M_{\Upsilon}} Q_\beta \bigg) +(\alpha \leftrightarrow \beta)
 \Bigg],
%--------------------------------------------------------------
\\
%--------------------------------------------------------------
%--------------------------------------------------------------
{\mathbb P}_{1,2}^{\mu\nu\alpha\beta} &=& {1\over 4 }
(g_{\perp\:\mu\nu} \, g_{\perp\:\alpha\beta}-g_{\perp\:\mu\alpha} \,
g_{\perp\:\nu\beta}- g_{\perp\:\mu\beta} \, g_{\perp\:\nu\alpha}).
%--------------------------------------------------------------
%--------------------------------------------------------------
\eqa
%--------------------------------------------------------------
\end{subequations}
%--------------------------------------------------------------
These five projection operators satisfy the normalization conditions
${\mathbb P}_{i;\mu\nu\alpha\beta}\,{\mathbb
P}_{j;}^{\mu\nu\alpha\beta}=\frac{1}{2}\delta_{ij}$ ($i$, $j$
corresponding to one of the five helicity configurations), except
that ${\mathbb P}_{0,0;\mu\nu\alpha\beta}\,{\mathbb
P}_{0,0}^{\mu\nu\alpha\beta}=1$.

All the helicity projectors are derived by obeying the exact decay
kinematics. Nevertheless, in this work we only target at the LO
accuracy in $v$ expansion. Therefore, Upon applying these projectors
to infer the intended helicity amplitudes, it is eligible to make
the following substitutions for the various quarkonium masses:
$M_{\Upsilon(nS)}\approx 2 m_b$ ($n=1,2,3$), and $M_{J/\psi}\approx
M_{\chi_{c0,1,2}}\approx 2m_c$. Implementing these approximations
considerably simplifies the corresponding loop calculation.

%--------------------------------------------------------------

%--------------------------------------------------------------

\begin{thebibliography}{99}
%--------------------------------------------------------------



%\cite{Beringer:1900zz}
\bibitem{Beringer:1900zz}
  J.~Beringer {\it et al.}  [Particle Data Group Collaboration],
  %``Review of Particle Physics (RPP),''
  Phys.\ Rev.\ D {\bf 86}, 010001 (2012).
  %%CITATION = PHRVA,D86,010001;%%

%\cite{Shen:2012iq}
\bibitem{Shen:2012iq}
  C.~P.~Shen {\it et al.}  [Belle Collaboration],
 % ``First observation of exclusive $\Upsilon(1S)$ and $\Upsilon(2S)$ decays into light hadrons,''
  Phys.\ Rev.\ D {\bf 86}, 031102 (2012)  [arXiv:1205.1246 [hep-ex]].
  %%CITATION = ARXIV:1205.1246;%%

%\cite{Jia:2006rx}
\bibitem{Jia:2006rx}
  Y.~Jia,
  %``Which hadronic decay modes are good for eta(b) searching: double J/psi or
  %something else?,''
  Phys.\ Rev.\  D {\bf 78}, 054003 (2008)
  [arXiv:hep-ph/0611130].
  %%CITATION = PHRVA,D78,054003;%%

%\cite{Gong:2008ue}
\bibitem{Gong:2008ue}
  B.~Gong, Y.~Jia and J.~X.~Wang,
  %``Exclusive eta(b) decay to double J / psi at next-to-leading order in
  %alpha(s),''
  Phys.\ Lett.\  B {\bf 670} (2009) 350
  [arXiv:0808.1034 [hep-ph]].
  %%CITATION = PHLTA,B670,350;%%

%\cite{Braguta:2009xu}
\bibitem{Braguta:2009xu}
  V.~V.~Braguta and V.~G.~Kartvelishvili,
  %``Decay eta(b) ---> J/psi J/psi in light cone formalism,''
  Phys.\ Rev.\  D {\bf 81} (2010) 014012
  [arXiv:0907.2772 [hep-ph]].
  %%CITATION = PHRVA,D81,014012;%%

%\cite{Sun:2010qx}
\bibitem{Sun:2010qx}
  P.~Sun, G.~Hao and C.~F.~Qiao,
  %``Pseudoscalar Quarkonium Exclusive Decays to Vector Meson Pair,''
  Phys.\ Lett.\  B {\bf 702}, 49 (2011)
  [arXiv:1005.5535 [hep-ph]].
  %%CITATION = PHLTA,B702,49;%%

%\cite{Braguta:2005gw}
\bibitem{Braguta:2005gw}
  V.~V.~Braguta, A.~K.~Likhoded and A.~V.~Luchinsky,
  %``Observation potential for chi/b at the Tevatron and LHC,''
  Phys.\ Rev.\ D {\bf 72}, 094018 (2005)
  [arXiv:hep-ph/0506009].
  %%CITATION = HEP-PH 0506009;%%

%\cite{Braguta:2009df}
\bibitem{Braguta:2009df}
  V.~V.~Braguta, A.~K.~Likhoded and A.~V.~Luchinsky,
  %``Double charmonium production in exclusive bottomonia decays,''
  Phys.\ Rev.\ D {\bf 80}, 094008 (2009)  [Erratum-ibid.\ D {\bf 85}, 119901 (2012)]
  [arXiv:0902.0459 [hep-ph]].  %%CITATION = ARXIV:0902.0459;%%

%\cite{Zhang:2011ng}
\bibitem{Zhang:2011ng}
  J.~Zhang, H.~Dong and F.~Feng,
  %``Exclusive decay of P-wave Bottomonium into double J/\psi,''
  Phys.\ Rev.\ D {\bf 84}, 094031 (2011)  [arXiv:1108.0890 [hep-ph]].
  %%CITATION = ARXIV:1108.0890;%%


%\cite{Sang:2011fw}
\bibitem{Sang:2011fw}
  W.~-L.~Sang, R.~Rashidin, U-R.~Kim and J.~Lee,
  %``Relativistic Corrections to the Exclusive Decays of C-even Bottomonia into S-wave Charmonium Pairs,''
  Phys.\ Rev.\ D {\bf 84}, 074026 (2011)  [arXiv:1108.4104 [hep-ph]].
  %%CITATION = ARXIV:1108.4104;%%

%\cite{Chen:2012ih}
\bibitem{Chen:2012ih}
  L.~-B.~Chen and C.~-F.~Qiao,
  %``P-wave Quarkonium Decays to Meson Pair,''
  JHEP {\bf 1211}, 168 (2012)  [arXiv:1204.0215 [hep-ph]].
  %%CITATION = ARXIV:1204.0215;%%

%\cite{Irwin:1990fn}
\bibitem{Irwin:1990fn}
  B.~A.~Irwin, B.~Margolis and H.~D.~Trottier,
  %``Vector to vector plus pseudoscalar decay modes in perturbative QCD,''
  Phys.\ Rev.\ D {\bf 42}, 1577 (1990).  %%CITATION = PHRVA,D42,1577;%%

%\cite{Jia:2007hy}
\bibitem{Jia:2007hy}
  Y.~Jia,
  %``Exclusive Double Charmonium Production from Upsilon Decay,''
  Phys.\ Rev.\  D {\bf 76} (2007) 074007
  [arXiv:0706.3685 [hep-ph]].
  %%CITATION = PHRVA,D76,074007;%%



%\cite{Abe:2002rb}
\bibitem{Abe:2002rb}
  K.~Abe {\it et al.}  [Belle Collaboration],
  Phys.\ Rev.\ Lett.\  {\bf 89}, 142001 (2002),
  [arXiv:hep-ex/0205104].

%\cite{Abe:2004ww}
\bibitem{Abe:2004ww}
  K.~Abe {\it et al.}  [Belle Collaboration],
  %``Study of double charmonium production in e+ e- annihilation at s**(1/2)
  %approx. 10.6-GeV,''
  Phys.\ Rev.\  D {\bf 70}, 071102 (2004)
  [arXiv:hep-ex/0407009].
  %%CITATION = PHRVA,D70,071102;%%

%\cite{Aubert:2005tj}
\bibitem{Aubert:2005tj}
  B.~Aubert {\it et al.}  [BABAR Collaboration],
  %``Measurement of double charmonium production in $e^+e^-$ annihilations at
  %$\sqrt{s}=10.6$ GeV,''
  Phys.\ Rev.\  D {\bf 72}, 031101 (2005)
  [arXiv:hep-ex/0506062].
  %%CITATION = PHRVA,D72,031101;%%

%\cite{Braaten:2002fi}
\bibitem{Braaten:2002fi}
  E.~Braaten and J.~Lee,
  %``Exclusive double-charmonium production in e+ e- annihilation,''
  Phys.\ Rev.\ D {\bf 67}, 054007 (2003)
  [arXiv:hep-ph/0211085].

\bibitem{Liu:2002wq}
  K.~Y.~Liu, Z.~G.~He and K.~T.~Chao,
  %``Problems of double charm production in e+ e- annihilation at s**(1/2) =
  %10.6-GeV. ((V)),''
  Phys.\ Lett.\  B {\bf 557}, 45 (2003)
  [arXiv:hep-ph/0211181].
  %%CITATION = PHLTA,B557,45;%%

%\cite{Hagiwara:2003cw}
\bibitem{Hagiwara:2003cw}
  K.~Hagiwara, E.~Kou and C.~F.~Qiao,
  %``Exclusive $J/\psi$ productions at $e^{+} e^{-}$ colliders,''
  Phys.\ Lett.\  B {\bf 570} (2003) 39
  [arXiv:hep-ph/0305102].
  %%CITATION = PHLTA,B570,39;%%

%\cite{Ma:2004qf}
\bibitem{Ma:2004qf}
  J.~P.~Ma and Z.~G.~Si,
  %``Predictions for e+ e- --> J/psi eta/c with light-cone wave-functions,''
  Phys.\ Rev.\  D {\bf 70}, 074007 (2004)
  [arXiv:hep-ph/0405111].
  %%CITATION = PHRVA,D70,074007;%%


%\cite{Bondar:2004sv}
\bibitem{Bondar:2004sv}
  A.~E.~Bondar and V.~L.~Chernyak,
  %``Is the BELLE result for the cross section sigma(e+ e- --> J/psi +  eta/c) a
  %real difficulty for QCD?,''
  Phys.\ Lett.\  B {\bf 612}, 215 (2005)
  [arXiv:hep-ph/0412335].
  %%CITATION = PHLTA,B612,215;%%

%\cite{Bodwin:2006dm}
\bibitem{Bodwin:2006dm}
  G.~T.~Bodwin, D.~Kang and J.~Lee,
  %``Reconciling the light-cone and NRQCD approaches to calculating e+ e- -->
  %J/psi + eta/c,''
  Phys.\ Rev.\  D {\bf 74}, 114028 (2006)
  [arXiv:hep-ph/0603185].
  %%CITATION = PHRVA,D74,114028;%%


%\cite{Zhang:2005cha}
\bibitem{Zhang:2005cha}
  Y.~-J.~Zhang, Y.~-j.~Gao, K.~-T.~Chao,
  %``Next-to-leading order QCD correction to e+ e- ---> J / psi + eta(c) at s**(1/2) = 10.6-GeV,''
  Phys.\ Rev.\ Lett.\  {\bf 96}, 092001 (2006).
  [hep-ph/0506076].

\bibitem{Gong:2007db}
  B.~Gong and J.~X.~Wang,
  %``QCD corrections to J/psi plus eta_c production in e+e- annihilation at
  %sqrt{s}=10.6 GeV,''
  Phys.\ Rev.\  D {\bf 77}, 054028 (2008)
  [arXiv:0712.4220 [hep-ph]].
  %%CITATION = PHRVA,D77,054028;%%

%\cite{He:2007te}
\bibitem{He:2007te}
  Z.~G.~He, Y.~Fan and K.~T.~Chao,
  %``Relativistic corrections to $J/\psi$ exclusive and inclusive double   charm
  %production at B factories,''
  Phys.\ Rev.\  D {\bf 75}, 074011 (2007)
  [arXiv:hep-ph/0702239].
  %%CITATION = PHRVA,D75,074011;%%

%\cite{Bodwin:2007ga}
\bibitem{Bodwin:2007ga}
  G.~T.~Bodwin, J.~Lee and C.~Yu,
  %``Resummation of Relativistic Corrections to e+ e- -> J/psi+eta_c,''
  Phys.\ Rev.\  D {\bf 77}, 094018 (2008)
  [arXiv:0710.0995 [hep-ph]].
  %%CITATION = PHRVA,D77,094018;%%

%\cite{Braguta:2008tg}
\bibitem{Braguta:2008tg}
  V.~V.~Braguta,
  %``Double charmonium production at B-factories within light cone formalism,''
  Phys.\ Rev.\ D {\bf 79}, 074018 (2009)
  [arXiv:0811.2640 [hep-ph]].  %%CITATION = ARXIV:0811.2640;%%

%\cite{Brambilla:2010cs}
\bibitem{Brambilla:2010cs}
 For a recent review on $e^+e^-\to J/\psi+\eta_c$ at $B$ factory,
 see N.~Brambilla {\it et al.},
  %``Heavy quarkonium: progress, puzzles, and opportunities,''
  Eur.\ Phys.\ J.\  C {\bf 71}, 1534 (2011).
 % [arXiv:1010.5827 [hep-ph]].
  %%CITATION = EPHJA,C71,1534;%%

%\cite{Dong:2012xx}
\bibitem{Dong:2012xx}
  H.~-R.~Dong, F.~Feng and Y.~Jia,
  %``$O(\alpha_s v^2)$ correction to $e^+e^-\to J/\psi+\eta_c$ at $B$ factories,''
  Phys.\ Rev.\ D {\bf 85}, 114018 (2012)  [arXiv:1204.4128 [hep-ph]].
  %%CITATION = ARXIV:1204.4128;%%

%\cite{Zhang:2008gp}
\bibitem{Zhang:2008gp}
  Y.~-J.~Zhang, Y.~-Q.~Ma and K.~-T.~Chao,
  %``Factorization and NLO QCD correction in e+ e- ---> J / psi (psi(2S)) + chi(c0) at B Factories,''
  Phys.\ Rev.\ D {\bf 78}, 054006 (2008)  [arXiv:0802.3655 [hep-ph]].
  %%CITATION = ARXIV:0802.3655;%%


%\cite{Wang:2011qg}
\bibitem{Wang:2011qg}
  K.~Wang, Y.~-Q.~Ma, K.~-T.~Chao,
  %``QCD corrections to e^+ e^- to J/\psi(\psi(2S))+\chi_{cJ} (J=0,1,2) at B Factories,''
  Phys.\ Rev.\  {\bf D84}, 034022 (2011).
%  [arXiv:1107.2646 [hep-ph]].

%\cite{Dong:2011fb}
\bibitem{Dong:2011fb}
  H.~R.~Dong, F.~Feng and Y.~Jia,
  %``$O(\alpha_s)$ corrections to $J/\psi+\chi_{cJ}$ production at $B$
  %factories,''
  JHEP {\bf 1110} (2011) 141.
  %[arXiv:1107.4351 [hep-ph]].
  %%CITATION = JHEPA,1110,141;%%


%\cite{Bodwin:1994jh}
\bibitem{Bodwin:1994jh}
  G.~T.~Bodwin, E.~Braaten and G.~P.~Lepage,
  %``Rigorous QCD analysis of inclusive annihilation and production of heavy
  %quarkonium,''
  Phys.\ Rev.\  D {\bf 51}, 1125 (1995)
  [Erratum-ibid.\  D {\bf 55}, 5853 (1997)]
  [arXiv:hep-ph/9407339].
  %%CITATION = PHRVA,D51,1125;%%


%\cite{Xu:2011:thesis}
\bibitem{Xu:2011:thesis}
  J.~Xu, Ph.D. thesis, {\it Electromagnetic transitions of quarkonia and
  productions of $P$-wave quarkonia}, IHEP (2011) (in Chinese).



%\cite{LopezCastro:1994xw}
\bibitem{LopezCastro:1994xw}
  G.~Lopez Castro, J.~L.~Lucio M. and J.~Pestieau,
  AIP\ Conf.\ Proc. {\bf 342}, 441 (1995)
[arXiv:hep-ph/9902300].

%\cite{Baldini:1998en}
\bibitem{Baldini:1998en}
  R.~Baldini {\it et al.},
  %``Measurement of J/psi --> N anti-N branching ratios and estimate of the
  %phase of the strong decay amplitude,''
  Phys.\ Lett.\  B {\bf 444}, 111 (1998).
  %%CITATION = PHLTA,B444,111;%%

%\cite{Suzuki:1999nb}
\bibitem{Suzuki:1999nb}
  M.~Suzuki,
  %``A large final-state interaction in the 0- 0- decays of J/psi,''
  Phys.\ Rev.\  D {\bf 60}, 051501 (1999)
  [arXiv:hep-ph/9901327].
  %%CITATION = PHRVA,D60,051501;%%


%\cite{Wang:2003hy}
\bibitem{Wang:2003hy}
  P.~Wang, C.~Z.~Yuan and X.~H.~Mo,
  %``Possible large phase in psi(2S) --> 1- 0- decays,''
  Phys.\ Rev.\  D {\bf 69}, 057502 (2004)
  [arXiv:hep-ph/0303144].
  %%CITATION = PHRVA,D69,057502;%%


%\cite{Yuan:2003hj}
\bibitem{Yuan:2003hj}
  C.~Z.~Yuan, P.~Wang and X.~H.~Mo,
  %``Relative phase between strong and electromagnetic amplitudes in psi(2S)
  %--> 0- 0- decays,''
  Phys.\ Lett.\  B {\bf 567}, 73 (2003)
  [arXiv:hep-ph/0305259].
  %%CITATION = PHLTA,B567,73;%%


%\cite{Dobbs:2006fj}
\bibitem{Dobbs:2006fj}
  S.~Dobbs {\it et al.}  [CLEO Collaboration],
  %``Measurement of interference between electromagnetic and strong  amplitudes
  %in psi(2S) decays to two pseudoscalar mesons,''
  Phys.\ Rev.\  D {\bf 74}, 011105 (2006)
  [arXiv:hep-ex/0603020].
  %%CITATION = PHRVA,D74,011105;%%


%\cite{Jacob:1959at}
\bibitem{Jacob:1959at}
  M.~Jacob and G.~C.~Wick,
  %``On the general theory of collisions for particles with spin,''
  Annals Phys.\  {\bf 7} (1959) 404
  [Annals Phys.\  {\bf 281} (2000) 774].
  %%CITATION = APNYA,281,774;%%

\cite{Haber:1994pe}
\bibitem{Haber:1994pe}
  H.~E.~Haber,
  %``Spin formalism and applications to new physics searches,''
  In *Stanford 1993, Spin structure in high energy processes* 231-272  [hep-ph/9405376].
  %%CITATION = HEP-PH/9405376;%%

%\cite{Brodsky:1981kj}
\bibitem{Brodsky:1981kj}
  S.~J.~Brodsky and G.~P.~Lepage,
  %``Helicity Selection Rules And Tests Of Gluon Spin In Exclusive QCD
  %Processes,''
  Phys.\ Rev.\  D {\bf 24}, 2848 (1981).
  %%CITATION = PHRVA,D24,2848;%%


%\cite{Lepage:1980fj}
\bibitem{Lepage:1980fj}
  G.~P.~Lepage and S.~J.~Brodsky,
  %``Exclusive Processes in Perturbative Quantum Chromodynamics,''
  Phys.\ Rev.\ D {\bf 22}, 2157 (1980).  %%CITATION = PHRVA,D22,2157;%%


%\cite{Chernyak:1983ej}
\bibitem{Chernyak:1983ej}
  V.~L.~Chernyak and A.~R.~Zhitnitsky,
  %``Asymptotic Behavior Of Exclusive Processes In QCD,''
  Phys.\ Rept.\  {\bf 112}, 173 (1984);
  %%CITATION = PRPLC,112,173;%%

%\cite{Bodwin:2008nf}
\bibitem{Bodwin:2008nf}
  G.~T.~Bodwin, X.~Garcia i Tormo and J.~Lee,
  %``Factorization theorems for exclusive heavy-quarkonium production,''
  Phys.\ Rev.\ Lett.\  {\bf 101}, 102002 (2008)  [arXiv:0805.3876 [hep-ph]].
  %%CITATION = ARXIV:0805.3876;%%


%\cite{Hahn:1998yk}
\bibitem{Hahn:1998yk}
  T.~Hahn and M.~Perez-Victoria,
  %``Automatized one loop calculations in four-dimensions and D-dimensions,''
  Comput.\ Phys.\ Commun.\  {\bf 118}, 153 (1999).
%  [arXiv:hep-ph/9807565].
  %%CITATION = CPHCB,118,153;%%


%\cite{Smirnov:2008iw}
\bibitem{Smirnov:2008iw}
  A.~V.~Smirnov,
  %``Algorithm FIRE -- Feynman Integral REduction,''
  JHEP {\bf 0810}, 107 (2008).
  [arXiv:0807.3243 [hep-ph]].

%\cite{Feng:2012iq}
\bibitem{Feng:2012iq}
  F.~Feng,
  %``$Apart: A Generalized Mathematica Apart Function,''
  Comput.\ Phys.\ Commun.\  {\bf 183}, 2158 (2012)  [arXiv:1204.2314 [hep-ph]].
  %%CITATION = ARXIV:1204.2314;%%


%\cite{Jia:2010fw}
\bibitem{Jia:2010fw}
  Y.~Jia, J.~X.~Wang and D.~Yang,
  %``Bridging light-cone and NRQCD approaches: asymptotic behavior of $B_c$
  %electromagnetic form factor,''
  JHEP {\bf 1110}, 105 (2011).
% [arXiv:1012.6007 [hep-ph]].
  %%CITATION = JHEPA,1110,105;%%

%\cite{Ellis:2007qk}
\bibitem{Ellis:2007qk}
  R.~K.~Ellis and G.~Zanderighi,
  %``Scalar one-loop integrals for QCD,''
  JHEP {\bf 0802}, 002 (2008)
  [arXiv:0712.1851 [hep-ph]].
Also see the website http://qcdloop.fnal.gov.
  %%CITATION = ARXIV:0712.1851;%%


%\cite{Eichten:1995}
\bibitem{Eichten:1995}
  E.~J.~Eichten and C.~Quigg,
  %``J/psi Decays,''
  Phys.\ Rev.\ D  {\bf 52}, 1726 (1995).
  %%CITATION = PRPLC,174,67;%%

%\cite{Gerard:1999uf}
\bibitem{Gerard:1999uf}
  J.~M.~Gerard and J.~Weyers,
  %``Phases and amplitudes in inclusive psi and psi' decays,''
  Phys.\ Lett.\  B {\bf 462}, 324 (1999)
  [arXiv:hep-ph/9906357].

%\cite{Shen:2012ei}
\bibitem{Shen:2012ei}
  C.~P.~Shen {\it et al.}  [Belle Collaboration],
  %``Search for double charmonium decays of the P-wave spin-triplet bottomonium states,''
  Phys.\ Rev.\ D {\bf 85}, 071102 (2012)  [arXiv:1203.0368 [hep-ex]].
  %%CITATION = ARXIV:1203.0368;%%

%\cite{Shen:2012:private}
\bibitem{Shen:2012:private}
  C.~P.~Shen, private communication.

%--------------------------------------------------------------
\end{thebibliography}
\end{document}